\documentclass[fleqn,10pt]{wlscirep}

\usepackage[utf8]{inputenc}
\usepackage[T1]{fontenc}
\usepackage{amsfonts}
\usepackage[]{algorithm2e}
\usepackage{comment}
\usepackage{graphicx}
\usepackage{hyperref}
\usepackage{breakurl}
\usepackage{tabularx,ragged2e}
\usepackage{bbold}
\usepackage{multirow}
\usepackage{bm}
\usepackage{color, colortbl}

\definecolor{Gray}{gray}{0.9}

\DeclareMathOperator*{\argmin}{arg\,min}

\newcommand{\beginsupplement}{
\setcounter{table}{0}
\renewcommand{\thetable}{S\arabic{table}}
\setcounter{figure}{0}
\renewcommand{\thefigure}{S\arabic{figure}}
\setcounter{equation}{0}
\renewcommand{\theequation}{S\arabic{equation}}
\setcounter{algocf}{0}
\renewcommand{\thealgocf}{S\arabic{algocf}}
}

\usepackage{xcolor}
\usepackage{caption}

\title{Robust estimation of SARS-CoV-2 epidemic { in} US counties}

\author[1,+]{Hanmo Li}
\author[1,*,+]{Mengyang Gu}
\affil[1]{Department of Statistics and Applied Probability, University of California, Santa Barbara, California, USA, 93106}
\affil[*]{Corresponding author. Email: \href{mailto:mengyang@pstat.ucsb.edu}{mengyang@pstat.ucsb.edu}}
\affil[+]{the authors contributed equally to this work}

\begin{abstract}

The COVID-19 outbreak is asynchronous { in} US counties. Mitigating the COVID-19 transmission requires not only the state and federal level order of protective measures such as social distancing and testing, but also public awareness of  time-dependent risk and reactions at county and community levels. We propose a robust approach to estimate the heterogeneous progression of SARS-CoV-2 at all US counties having no less than 2 COVID-19 associated deaths, and we { use} the daily probability of contracting (PoC) SARS-CoV-2 for a susceptible individual to quantify  the risk of SARS-CoV-2 transmission in {a} community. We found that shortening {by} $5\%$ of the infectious period of SARS-CoV-2 can reduce around $39\%$ (or $78$K, $95\%$ CI: $[66$K $, 89$K $]$) of the COVID-19  associated deaths in the US as of 20 September 2020.  Our findings also indicate that {reducing} infection and deaths by {a} shortened infectious period is more pronounced for areas with the effective reproduction number close to 1, suggesting that testing should be used along with other mitigation measures, such as social distancing and facial mask-wearing, to reduce the transmission rate. Our deliverable includes a dynamic county-level map for local officials to determine optimal policy responses and for {the} public to better understand the risk of contracting SARS-CoV-2 on each day. 

\end{abstract}
\begin{document}
 
\flushbottom
\maketitle
%
%
\thispagestyle{empty}


\section*{Introduction}
The outbreak of new coronavirus 2019 (COVID-19) has caused nearly 200,000 deaths in the US, and among those, there are 2,277 counties with no less than 2 associated deaths as of 20 September 2020 \cite{dong2020interactive}. The ongoing COVID-19 pandemic has led to unprecedented non-pharmaceutical interventions (NPIs), including travel restrictions, lockdowns, social distancing,  facial masks wearing, and quarantine to reduce the spread of SARS-CoV-2 in the US. 
The COVID-19 outbreak is prolonged and asynchronous across regions. 
{Thus it is critical} to estimate the dynamics of COVID-19 epidemic to determine appropriate protective measures before the availability of effective vaccines. 


A non-negligible proportion of SARS-CoV-2 infectious individuals is asymptomatic or have mild symptoms  \cite{li2020substantial}.
 We term the individuals the \textit{active infectious individuals} who can transmit the disease to others but may not be diagnosed yet. Identifying the number of {active infectious individuals}  is crucial to monitor the transmission in a community. Another important time-dependent quantity is the expected number of secondary cases resulted from each active infectious individual, or \textit{effective reproduction number}.  In this article, we estimate these two time-dependent quantities for all US counties with no less than 2 COVID-19 associated deaths as of {20 September 2020}; the population of some counties that falls within this category is even less than ten thousand. Furthermore, based on these two time-dependent quantities, a more interpretable measure, called the daily \textit{probability of contracting} (PoC) SARS-CoV-2 for an individual at {the} county-level was used to quantify the risk. {This static risk factor with fixed transmission rates was studied before \cite{hethcote2000mathematics}. Here we studied the dynamic  transmission rate parameter, which is estimated  by the number of deaths, test positive rates and the number of confirmed cases in a  community. The risk factor can be extended to measure the  risk of an event with different sizes \cite{chande2020real}.  } The fine-grain estimation of disease progression characteristics allows {the} public to understand the risk of contracting COVID-19 on a daily basis. 
 
Predictive mathematical models are useful for analyzing an epidemic to guide policy responses \cite{holmdahl2020wrong}. The epidemiology compartmental models such as SIR, SEIR, SIRD, and their extensions \cite{lin2020conceptual,giordano2020modelling,Dehningeabb9789,fernandez2020estimating,swan2020vaccines}, stochastic agent based models \cite{flaxman2020estimating,hoertel2020stochastic}, branching processes \cite{bertozzi2020challenges}, and network analysis \cite{firth2020using} have  advanced our understanding of transmission rates and incubation period of SARS-CoV-2,  which are connected to the traffic flow and mobility during the COVID-19 outbreaks at different regions \cite{jia2020population,badr2020association}. The disease progression characteristics, such as the transmission rate, are often estimated based on the daily death toll \cite{lin2020conceptual,fernandez2020estimating, flaxman2020estimating, hoertel2020stochastic}. 
 {However, it is challenging to estimate the progression of the epidemic in US counties with small population, because the number of daily observed confirmed cases and COVID-19-related deaths is small.} 
 Meanwhile, {using observed laboratory-confirmed COVID-19 cases (henceforth, observed confirmed cases)} might significantly underestimate the population that have been infected with the SARS-CoV-2. It was found in  \cite{anand2020prevalence} that around $9.3\%$ of the US individuals (or roughly 30 million)  may have contracted the COVID by July 2020 based on serology tests, whereas less than 4.8 million COVID-19 positive cases have been confirmed in the US {before} August 2020  \cite{dong2020interactive}. {Thus,} {it is important to estimate} the number of individuals who contracted COVID-19 but {had} not  tested positive. The focus herein is on integrating COVID-19-related death toll and test data to obtain a robust estimation of the disease progression characteristics of COVID-19 at county and community levels.

\begin{figure}[t]
\centering


\includegraphics[height=0.67\textwidth,width=0.8\textwidth ]{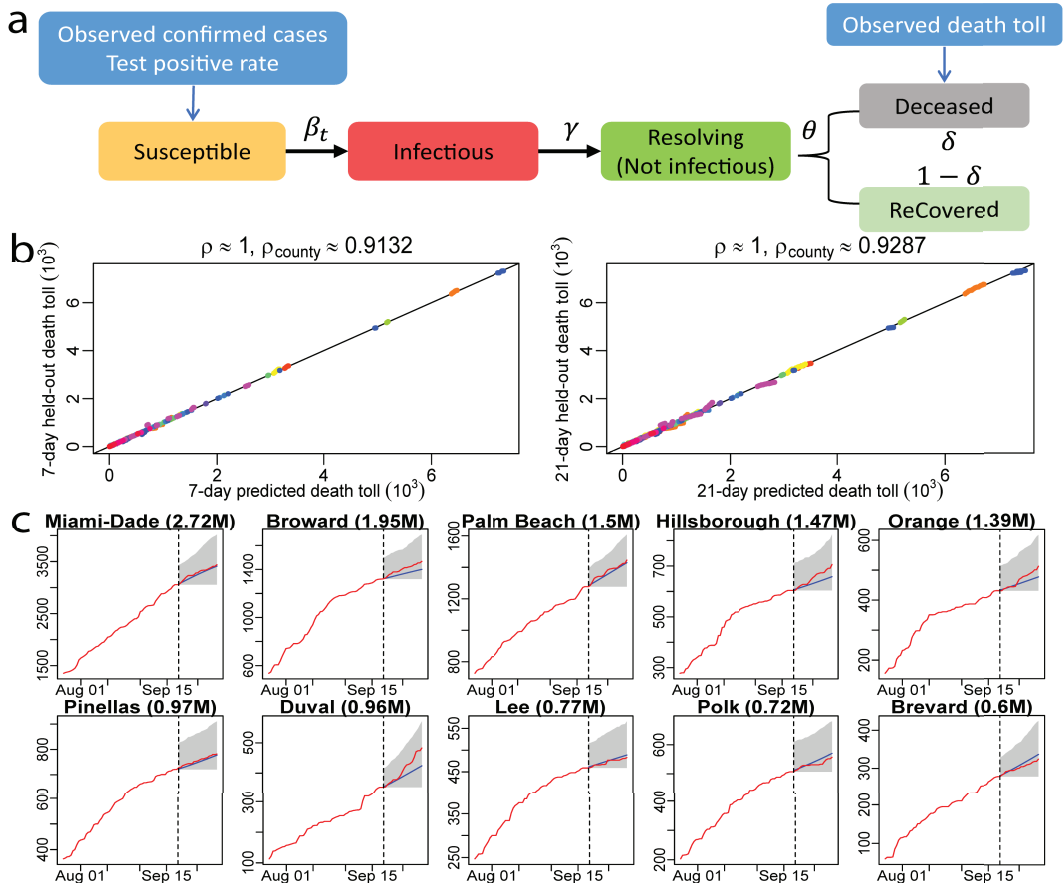}

\caption{a, The SIRDC model and the data used for analysis. b, 7-day death toll forecast and 21-day death toll forecast against the held-out truth in 2,277 US counties with no less than 2 deaths as of 20 September 2020. Each dot is a cumulative death toll for one county at one held-out day.  Counties from the same state are graphed using the same color. The Pearson correlation coefficient ($\rho$) of the nation and the weighted average of Pearson correlation coefficient for counties ($\rho_{county}$) are recorded.  c, 21-day death toll forecasts in 10 counties with largest population in Florida, where the red line represents the observed death toll and blue line means the forecast. The forecast starts from 21 September 2020, marked by the vertical black dash line. The grey shadow area is the 95$\%$ confidence interval of the forecast. Numbers in the parentheses right after the county name are population in million. The Extended Data Figures \ref{fig:90-day predictions in FL counties} and \ref{fig:90-day predictions in CA counties} show 21-day death toll forecast for all counties in Florida and  California.
 }
\label{fig:SIRDC_prediction}
\end{figure}

\begin{table}[t]
\caption{Policy summary}
\centering
\begin{tabularx}{\textwidth}{lX}

\hline 
Background & 
The transmission of SARS-CoV-2 is heterogeneous and asynchronous in US counties. It is thus important to assess the risk before lifting or replacing any mitigation measure in the community. 
We have developed a novel approach to integrate test data and death toll to estimate {the} probability of contracting COVID-19, as well as the time-dependent transmission rate and  number of active infectious individuals at the county level in the US. 
\\
\hline
Main findings and limitations & 
National level order of protective measures reduces the transmission rate and active number of infectious individuals for most US counties in April, whereas the risk of contracting  SARS-CoV-2 rebounded between late June and early July, as the protective measures were relaxed. 
We found that when the infectious period of SARS-CoV-2  is shortened by $5\%$ and $10\%$, the number of deaths can be reduced from $199$K to $120$K ($95\%$ CI: $[109$K, $132$K $]$) and $80$K ($95\%$ CI: $[72$K, $89$K$]$) as of 20 September 2020,  respectively, when other protective measures were kept the same. The reduction of {the} infectious period can be achieved by extra testing in addition to  ongoing protective measures. Our model relies on the existing knowledge of the COVID-19 and model assumptions. Other information, such as demographic profiles, mobility, and serology test data, can be used to calibrate the model parameters and assumptions at the community level.  \\
\hline
Policy implications & Our model indicates that extra testings, along with the current NPIs, can significantly reduce the number of deaths associated with COVID-19. The estimated probability of contracting  COVID-19  can be used as an interpretable risk factor to guide community policy responses.  \\
\hline
\end{tabularx}

\label{tab:policy_summay}
\end{table}


One critical quantity to evaluate an infectious disease outbreak is the time-dependent transmission rate, based on which one can compute the basic reproduction number and the effective reproduction number of the disease. Various approaches were proposed to estimate this parameter.  The transmission rate was modeled as a decreasing function of the time in \cite{lin2020conceptual}, a function of NPIs in \cite{flaxman2020estimating} and a geometric Brownian motion in \cite{kucharski2020early}. Unlike the outbreak in China or other {countries in north-east Asia},  transmission rates of the COVID-19 progression in the US does not monotonically decrease due to the prolonged duration of the outbreak, and it is challenging to determine a suitable parametric form of this parameter in terms of time. In \cite{fernandez2020estimating}, the transmission rate parameter was related to the initial values of infectious cases, resolving cases, and up to two derivatives of the daily death toll. This method provides a flexible way to estimate the time-dependent transmission rate from the death toll and its derivatives, yet  unstable for {counties} with moderate or small population {sizes}, as numerical estimation of the daily death toll and its derivatives is often unstable.  


In this work, we propose a robust approach of integrating test data and death toll to estimate COVID-19 transmission characteristics by a Susceptible, Infectious, Resolving (but not infectious), Deceased,  and reCovered (SIRDC) model initially studied in \cite{fernandez2020estimating}. We illustrate that the transition between different stages of disease progression in the SIRDC model in part a of  Figure \ref{fig:SIRDC_prediction}. First, a part of the population is infected by active infectious individuals each day, depending on the transmission rate parameter ($\beta_t$). After $\gamma^{-1}$ {days}, an active infectious individual is expected to be no longer infectious, denoted by the resolving compartment, meaning that this individual will not transmit COVID-19 to others as a result of hospitalization or self-quarantine. We term the average length of an active infectious individual the \textit{infectious period}.   A resolving case is expected to be resolved (either recovered or deceased) after $\theta^{-1}$ {days}. The proportion of deaths from the number of resolved cases is controlled by the fatality rate parameter $\delta$. 


{Our approach has three  innovations. First, we solve the compartmental models using a midpoint rule with a step size of 1 day, as the confirmed cases and death toll are updated daily in most US counties, and this is discussed in the method section. Second, we combine test positive rates, confirmed cases and death toll to estimate the daily transmission rate parameter.} Our estimate of transmission rates and reproduction numbers is robust and accurate to reproduce the number of {the} death toll and other compartments for counties with medium to small population sizes (Figure \ref{fig:map_state_county} and  Extended data Figure \ref{fig:real_comparison_Euler}). The simulated studies also suggest that our approach is more robust than the solution in \cite{fernandez2020estimating}  (Extended Data Figure \ref{fig:simulated_comparison_Euler}), as our solution does not require estimating derivatives of the daily death toll. Only two parameters, the initial values of the number of active infectious individuals and the number of resolving cases, need to be estimated numerically for each county. {Then we can solve the time-dependent transmission rates and all other compartments subsequently.} Since only two parameters are estimated for each county, our estimation rarely depends on initial values we choose for the optimization. 
{Finally, we use a Gaussian process to model the residual between the observed death toll and that from the SIRDC model, leading to more accurate predictions and proper uncertainty quantification. A summary of the main findings, limitations, and policy implication are given in Table \ref{tab:policy_summay}.}

\section*{Results}
\begin{table}[t]
\caption{Interpretation of the daily PoC SARS-CoV-2 in a community.}
\centering

\begin{tabular}{|l|l|l|l|l|l|}
\hline 
Daily PoC SARS-CoV-2 & $<0.001\%$ &$0.001\%$ to $0.01\%$ & $0.01\%$ to $0.1\%$ & $0.1\%$ to $1\%$ & $>1\%$\\
 \hline
Risk & controllable & moderate & alarming  & strongly alarming & hazardous  \\   
\hline

\end{tabular}

\label{tab:interpretation_PoC}
\end{table}

\begin{figure}[t]
\centering
\vspace{-.2in}
\includegraphics[height=0.3\textwidth,width=1\textwidth ]{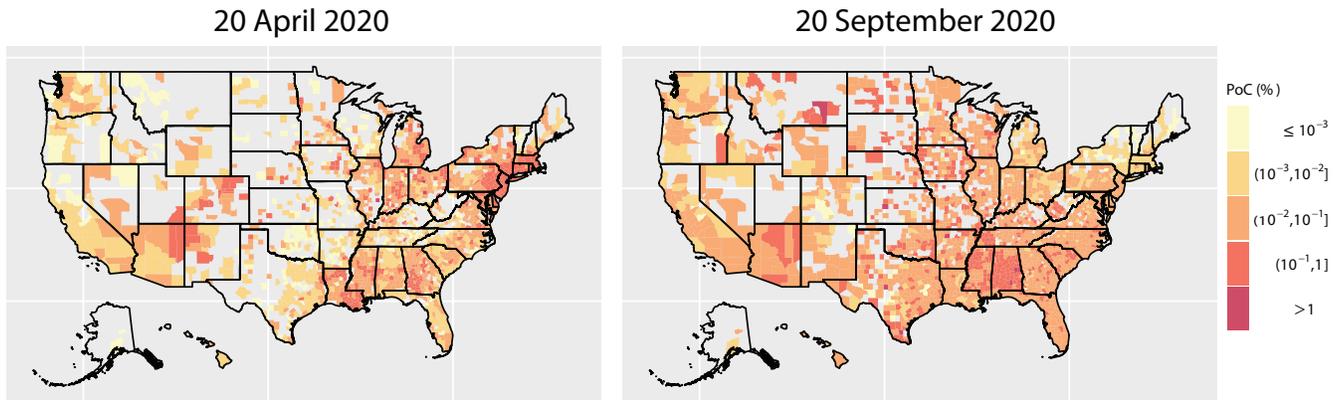}

\caption{a, The estimated probability of contracting SARS-CoV-2 at 1,856 counties on 2020-04-20, and b, at 2,277 counties on 20 September 2020.  The probability of contracting SARS-CoV-2 is truncated at $10^{-6}$, whereas only 78 counties on 20 April and 45 counties on 20 September are below this level, respectively. 
}
\label{fig:map_prob}
\end{figure}
\begin{figure}[!t]
\centering

\includegraphics[height=0.55\textwidth,width=1\textwidth ]{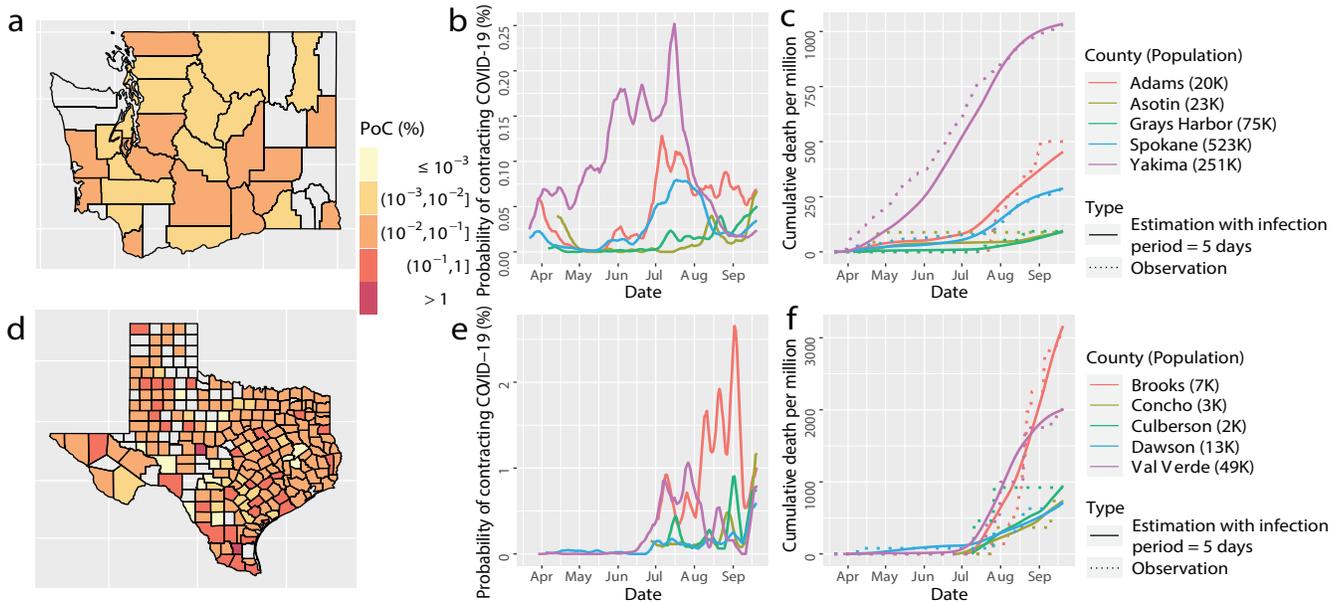}

\caption{a, The estimated probability of contracting SARS-CoV-2 in Washington state on 20 September 2020. b, the probability of contracting SARS-CoV-2 from  5 counties in Washington state with the largest {PoC SARS-CoV-2} values on 20 September 2020 . c, the observed (dots) and fitted (solid line) cumulative death toll in the 5 counties in figure b from the same time period. d-f, The results in Texas that have the same interpretation as a-c. Part e and f have different scales than part b and c, respectively.  
}
\label{fig:map_state_county}
\end{figure}

We first verify our model performance by forecasting at the county level. The 7-day and 21-day death projections for $2,277$ US counties using data by 20 September 2020, for instance, are close to the held-out test death toll in these counties, shown in part b and part c of Figure \ref{fig:SIRDC_prediction}. The Pearson correlation coefficient ($\rho$) is larger than 0.999  7-day and 21-day forecast. We also calculate the weighted average of Pearson correlation coefficient for counties ($\rho_{county}$), which treats each county as a different population and  population size is used to computed the weighted average of Pearson correlation coefficient for counties. 
The 21-day forecast of each considered county in Florida and California using observations by 20 September 2020 is provided in {Extended Data Figure \ref{fig:90-day predictions in FL counties} and \ref{fig:90-day predictions in CA counties}, respectively}. The death toll forecast based on our model is accurate for most US counties, and around $95\%$ of the held-out test data is covered by nominal $95\%$ predictive interval (Table S1 in supplementary materials), indicating that the uncertainty assessment is accurate. {To further test the predictive performance of our model, we use data by {1 December}, 2020 to make 21-day and 90-day predictions of deaths in the 10 largest counties in Florida and California. The forecast results are shown in Extended Data Figure \ref{fig:21-day prediction in ten largest CA and FL counties} and \ref{fig:90-day prediction in ten largest CA and FL counties}, respectively. While this is a challenging scenario, as confirmed cases and deaths increase dramatically across the US during the winter, we found that our 21-day predictions are reasonably accurate for all 20 counties. Thus, our models can be used reliably for the short-term projection of COVID-19 related deaths at the county level during different periods of the epidemic. Furthermore, a 90-day accurate forecast of US counties before the winter may be an almost impossible task, and indeed we underestimate  death counts for a few counties due to a rapid increase in death counts during the winter. On the other hand, our model that fuses test data and death toll correctly projects the rapid increase in death counts for most counties during the winter,  even if death counts do not increase dramatically during the training period. 
} 




Based on the robust estimation of transmission rates, we derived  the county-level estimation of daily PoC SARS-CoV-2. We classify the daily PoC SARS-CoV-2 in a community into five levels listed in Table \ref{tab:interpretation_PoC}. On 20 September 2020, out of 2,277 US counties, only 60 counties were at the controllable level and 311 counties were at the moderate level, whereas 1906 counties were at the either alarming, strongly alarming, or hazardous level. The daily PoC SARS-CoV-2 measures the average probability to contract SARS-CoV-2 for a susceptible individual in a community, and the risk varies from individuals to individuals. Nonetheless, the PoC SARS-CoV-2 is an interpretable measure for public understanding of the average risk of contracting SARS-CoV-2 in a community on a given day.


We graph the estimated  PoC SARS-CoV-2 of an individual at US counties on 20 April 2020 and 20 September 2020 in Figure \ref{fig:map_prob}. On 20 April 2020, the PoC SARS-CoV-2 is  large in northeastern regions and some southern states such as Arizona, New Mexico, and New Orleans. On 20 September 2020, the PoC SARS-CoV-2 is large in many inland states, for instance, Montana,  North Dakota, Mississippi, and Alabama. Although the PoC SARS-CoV-2 on 20 September in  northeastern regions is substantially lower than that on 20 April, the PoC SARS-CoV-2 for an individual is large in most other states on 20 September, suggesting that the relaxation of protective measures can lead to more population contracting COVID-19, and consequently more deaths at a rate no slower than that in late April.



\begin{figure}[t]
\centering

\includegraphics[height=0.3\textwidth,width=1\textwidth ]{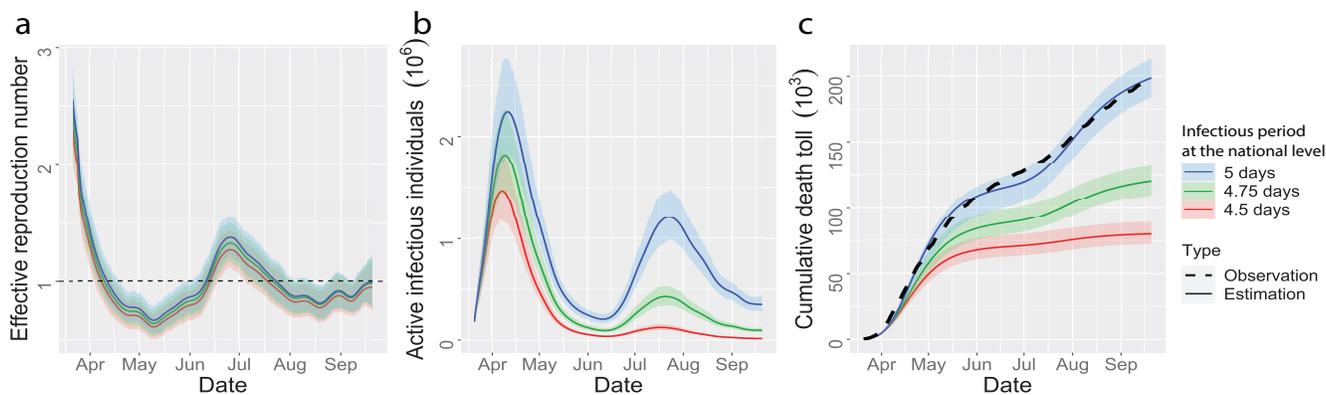}

\caption{a, b, The estimate reproduction number and overall number of active {infective individuals} in the US, including 50 states and Washington D.C., from 21 March 2020 to 20 September 2020 with infectious period assumed to be 5 days (blue), 4.75 days (green) and 4.5 days (red). c, The estimate overall death toll in the US. The time period and interpretation of c are aligned with a and b, except that the black dots in c stand for the observed death toll in the US. 
}
\label{fig:us_diff_infection_period}
\end{figure}

Officials can use the daily PoC SARS-CoV-2  to determine whether the  mitigation policies can be lifted or replaced by other measures for different regions. The probability of contracting COVID-19 in many counties in Texas on 20 September 2020, for example, is  larger than those in Washington (part (a) and (d) in Figure \ref{fig:map_state_county}), indicating that {Texas should undertake more protective measures} to reduce the risk. The nationwide lockdown order and social distancing in spring effectively reduced the PoC SARS-CoV-2 in 4 out of 5 counties in Washington, while the PoC SARS-CoV-2 of all counties increases in late June and early July, as some of the nonpharmaceutical interventions (NPIs) were lifted (part b in Figure \ref{fig:map_state_county}). Part (c) shows that the model {fits the} death toll. With only two parameters estimated numerically for each county, the fit is reasonably good for these counties  at a wide range of dates. In comparison, though the outbreak of 5 counties in Texas started in early summer, the PoC SARS-CoV-2 in these Texas counties is much higher than {that in Washington counties on 20 September}  (part (e) in Figure \ref{fig:map_state_county}).  Our model also fits the death toll  of the counties in Texas relatively well (part (f) in Figure \ref{fig:map_state_county}). 
The county-level estimation and forecast are updated regularly on the COVID-19 US dashboard: \href{https://covid19-study.pstat.ucsb.edu/}{https://covid19-study.pstat.ucsb.edu/}.

The effectiveness of protective measures were studied to reduce the transmission rate \cite{giordano2020modelling, Dehningeabb9789, flaxman2020estimating, hoertel2020stochastic, firth2020using, davies2020effects}, whereas the efficacy of these measures depends on the reactions from the public, which is likely to vary from region to region.  Another simultaneous effort to mitigate the spread of {the} COVID-19 outbreak is through testing and contact tracing, which reduces the infectious period, and consequently, the number of active infectious individuals. For Washington and Texas, we simulate the model output with infectious period reduced by $5\%$ (or equivalently 4.75 days in total), while the transmission rate ($\beta_t$ in SIRDC model) is held the same. We found that the PoC SARS-CoV-2 is reduced by 5 times for 12 counties out of 28 considered counties in Washington and 6 counties out of 209 considered counties in Texas, as shown in the {Extended Data Figure \ref{fig:map_state_county_with_infection_period_4.75}}.   Furthermore, when we reduce the infectious period by $10\%$ (or equivalently 4.5 days in total), while the transmission rate ($\beta_t$ in SIRDC model) is held the same, the PoC SARS-CoV-2 is reduced by 5 times for 26 {out of 28} counties in Washington and 146 {out of 209} counties in Texas, shown in {Extended Data Figure \ref{fig:map_state_county_with_infection_period_4.5}}. 






 We graph the estimated effective reproduction number, the number of active infectious individuals, and {the} cumulative death toll in the US, along with the simulated values when the average infectious period is reduced {from 5 days} to 4.75 days and 4.5 days in Figure \ref{fig:us_diff_infection_period}. First, we found that mitigation measures in March effectively reduce the effective reproduction number to below 1, whereas the value rebounded in summer after some of these measures were relaxed in different regions. Consequently, {the} US has experienced two waves of the outbreak in terms of the number of active infectious individuals (part (b) in Figure \ref{fig:us_diff_infection_period}). The high test positive rate at the beginning of the epidemic (Extended Data Figure 5) indicates that a substantial number of active infectious individuals  were not diagnosed in April due to the lack of diagnostic tests.  According to our estimates, the peak of the first wave in April is larger than that of the second wave in July in terms of the number of active infectious individuals, whereas the peak of the daily observed confirmed cases in April is smaller than that of the second wave in July (Extended Data Figure 5).
 
 Second,  the simulated results suggest that {shortening} infectious period of SARS-CoV-2 by $5\%$ and $10\%$ can reduce the total deaths from $199$K to $120$K ($95\%$ CI: $[109$K, $132$K$]$) and $80$K ($95\%$ CI: $[72$K, $89$K$]$), respectively,  as of 20 September 2020, when other protective measures were held as the same (part c in Figure \ref{fig:us_diff_infection_period}). Note that since we held the transmission rate parameter ($\beta_t$) to be the same (a scenario where the public adheres to the protective measure same as the reality), the effective reproduction number {barely changes} (part a in Figure \ref{fig:us_diff_infection_period}).  However, the slightly shortened infectious periods of SARS-CoV-2 can reduce the death toll substantially (part c in Figure \ref{fig:us_diff_infection_period}), as the  number of active infectious individuals decreases (part b in Figure \ref{fig:us_diff_infection_period}).

We found that a shortened infectious period substantially reduces the number of active infectious individuals and {fatalities} in the second wave. {However,} the changes are smaller in the first wave, since the effective reproduction number in the second wave is smaller than that in the first wave (Figure \ref{fig:us_diff_infection_period}). 
The county level estimation also validates this point (Extended Data Figure \ref{fig:map_state_county_with_infection_period_4.75} and \ref{fig:map_state_county_with_infection_period_4.5}). 
This finding indicates that the efforts of shortening the infectious period of SARS-CoV-2 should not replace the other protective measures, such as social distancing and facial mask-wearing to reduce the transmission rate.






Diagnostic tests can be used to shorten the length of {the} infectious period of an active infectious individual.  Drastically reducing the infectious period may not be possible without contact tracing, which is challenging when {there is a large number of active infective cases}. Reducing {the} infectious period by around $5\%$, in comparison,  {may} be achieved by periodically diagnostic tests every 20 days for each susceptible individual. {More frequent testing or contact tracing may be needed to achieve this goal, as the infection is most likely to happen between days 2 and 6 after exposure due to the high viral load of SARS-CoV-2  \cite{goyal2021viral}.} Another efficient way is to test susceptible individuals with a high risk of contracting or spreading SARS-CoV-2, such as individuals with more daily contacts or have contacts with vulnerable populations, e.g., workers from senior living facilities.   Our estimation of {the PoC} SARS-CoV-2 can be used as a response to develop regression models using covariates including demographic information and mobility to elicit personalized risk of contracting SARS-CoV-2 for susceptible individuals.

Finally, efforts on reducing the length of {the} infectious period should not replace other protective measures for reducing transmission rates of SARS-CoV-2, as the number of active infectious individuals and death toll can be effectively reduced {only if} the effective reproduction number is not substantially  larger than 1. 

\section*{Discussion}
Our study has several limitations. First, our findings are based on the  available knowledge and model assumptions, as with all other  studies.  One critical parameter is the death rate, assumed to be $0.66\%$ on average \cite{verity2020estimates}, whereas this parameter can vary across regions due to the demographic profile of the population and available medical resources. The studies of {the} prevalence of SARS-CoV-2 antibodies based on serology tests \cite{anand2020prevalence} can be used to determine the size of {the} population {who have} contracted SARS-CoV-2, and thus provides estimates on {the} death rate, as the death toll is observed. Besides, we assume the infected population can develop immunity since recovery for a few months, which is commonly used in other models. The exact duration of {immunity post-infection}, however, remains unverified scientifically. 
Third, we assume that the number of susceptible individuals and, consequently, the number of individuals {who} have contracted SARS-CoV-2 can be written as a function of the number of observed confirmed cases and test positive rates, calibrated based on the death toll.
More information such as the proportion of population adhere to the mitigation measures, mobility, and demographic profile  can be used to improve the estimation of susceptible individuals in a region. 

Our results can be used to mitigate the ongoing pandemic of  SARS-COV-2 and other infectious disease outbreaks in {the} future. The estimated daily PoC SARS-CoV-2 at the county level, for example, is an interpretable measure to understand the risk of contracting COVID-19 on a daily basis and a surveillance marker to determine appropriate policy responses. Besides, Our method can be extended when an effective vaccine becomes available \cite{swan2020vaccines}.  Finally,  further studies of this measure relative to different mobility, demographic information, and social-economic status can provide more precise guidance for local officials to protect vulnerable populations from contracting SARS-CoV-2, when an effective vaccine is not available. 


\begin{table}[]
\centering
\caption{{Main symbols and definitions in the Methods Section.}}
\label{tab: symbols in methods}
\begin{tabular}{|l|l|}
\hline
\rowcolor{Gray}
\textbf{Symbol}             & \textbf{Definition}                                            \\ \hline
$S(t)$             & number of susceptible cases on day t                 \\ \hline
$I(t)$             & number of infectious cases which can transmit COVID-19 on day t                       \\ \hline
$R(t)$             & number of resolved cases which get infected but cannot transmit COVID-19 on day t \\ \hline
$D(t)$             & number of deceased cases on day t                    \\ \hline
$C(t)$             & number of recovered cases on day t                   \\ \hline
$N$                & number of population in a given area                  \\ \hline
$\beta (t)$        & transmission rate on day t                           \\ \hline
${\gamma}^{-1}$ & average number of days an individual can transmit COVID-19                                           \\ \hline
${\theta}^{-1}$ & average number of days for a case to get resolved     \\ \hline
$\delta$           & proportion of deceased cases, a.k.a. fatality rate \\ \hline
$R_0(t)$           & basic reproduction number on day t                   \\ \hline
$R_{eff}(t)$       & effective reproduction number on day t               \\ \hline
$P(t)$             & average probability of contracting (PoC) SARS-CoV-2 on day t \\ \hline
$p(t)$             & state-level  test positive rate  on day t              \\ \hline
$c^o(t)$           & cumulative number of observed confirmed cases on day t          \\ \hline
$\Delta c^o(t)$    & daily number of observed confirmed cases on day t    \\ \hline
$c^u(t)$           & cumulative number of unobserved confirmed cases on day t        \\ \hline
$\alpha$           & power parameter for estimating the number of susceptible cases  \\ \hline
$\omega$           & weight parameter for estimating the number of susceptible cases \\ \hline
$z$                & zero-mean Gaussian process                            \\ \hline
\end{tabular}
\end{table}
\section*{Methods}
{We introduce our methods in this section. The main symbols used in this section and their definitions are provided in Table \ref{tab: symbols in methods}.}

{\bf SIRDC compartmental models}. The SIRDC model for the $j$th county in the $i$th state in the US is described below:
\begin{align}
\begin{split}
\dot{S}_{i,j}(t)&= \frac{-\beta_{i,j}(t) S_{i,j}(t) I_{i,j}(t)}{N_{i,j}}, \\ \dot{I}_{i,j}(t)&=\frac{\beta_{i,j}(t) S_{i,j}(t) I_{i,j}(t)}{ N_{i,j}}-{\gamma I_{i,j}(t)}, \\
\dot{R}_{i,j}(t)&= {\gamma I_{i,j}(t)}-{\theta R_{i,j}(t)}, \\ \dot{D}_{i,j}(t)&={\delta \theta R_{i,j}(t)},\\
\dot{C}_{i,j}(t)&={(1-\delta) \theta R_{i,j}(t)},
\end{split}
\label{equ:model_SIRDC}
\end{align}
 where $S_{i,j}(t)$, $I_{i,j}(t)$, $R_{i,j}(t)$,  $D_{i,j}(t)$ and $C_{i,j}(t)$ denote the number of individuals at these 5 compartmental groups on day $t$, respectively, and $N_{i,j}$ denotes the number of individuals in county $j$ {from} state $i$ for $i=1,2,...,k$, $j=1,2,...,n_{i}$ with $n_i$ being the number of counties of the $i$th state considered in the analysis and $t=1,2,...,T_{i,j}$. The time-dependent transmission rate parameter is denoted by $\beta_{i,j}(t)$ and  the inverse of average number of days an infectious individual can transmit the COVID-19 is denoted by $\gamma$. 
The inverse of the average number of dates for a case to get resolved (i.e. deceased or recovered)  is denoted by $\theta$ and the  proportion of deceased cases (i.e. death rate) is denoted by $\delta$. The parameters $(\gamma, \theta, \delta)$ were invariant over time and held fixed in this study. Following \cite{davies2020effects}, we assume the infectious period to be $5$ {days} on average, and a case is expected to resolve after $10$ days. The average death rate is assumed to be $0.66\%$ \cite{verity2020estimates}. Additional verification of these assumptions and sensitivity analysis of these parameters are provided in the supplementary materials.

 To determine the characteristics of the SARS-CoV-2 epidemic at US counties, we define the time-dependent \textit{effective reproduction number}, i.e. the average number of secondary cases per primary cases as  $\mathcal R_{eff}^{i,j}(t)=\mathcal R_{0}^{i,j}(t)S_{i,j}(t)/N_{i,j}$, where the $\mathcal R_{0}^{i,j}(t)=\beta_{i,j}(t)/\gamma$ denotes the \textit{basic reproduction number} on day $t$. When $\mathcal R_{eff}^{i,j}(t)<1$, it means that the number of the active infectious individuals will decrease (and vice versa, if $\mathcal R_{eff}^{i,j}(t)>1$). The effective reproduction number was often used to quantify whether or not the disease is under control \cite{nishiura2009effective}. {However}, the effective reproduction number does not directly quantify risk of contracting SARS-COV-2 for a susceptible individual, as the number of active infectious individuals in a region was not taken into consideration. We compute the average probability of contracting (PoC) SARS-CoV-2, denoted as $P_{i,j}(t)={\mathcal R^{i,j}_{eff}(t)I_{i,j}(t)}\gamma/({S_{i,j}(t)})={\mathcal \beta_{i,j}(t)I_{i,j}(t)}/{N_{i,j}}$, which  quantifies the risk of a susceptible individual in county $j$ from state $i$ to catch SARS-CoV-2  on day $t$. Here the risk is on an average sense among all susceptible individuals in a region.
 
 
 

The most critical parameter of the SIRDC model is the transmission rate parameter, $\beta_{i,j}(t)$, as a function of time, based on which we obtain the reproduction number on day $t$. To estimate the {time-dependent} transmission rates for communities with small population sizes, we derive  a more robust estimation of {the} transmission rate of each county based on {the} death toll and testing data, discussed below.


{\bf Closed-form expressions of the time-dependent transmission rates}. Since the observations such as death toll and confirmed cases are {generally} updated daily, we {solve the ordinary differential equations (ODEs) in the SIRDC model (equation (\ref{equ:model_SIRDC}))  approximately} by the midpoint rule of the integral with a step size of $1$ day. For day $t \in \mathbb{N}^{+}$, the approximation is described below:
\begin{align}
\frac{S_{i,j}(t+1)}{S_{i,j}(t)} \doteq&~\exp\left\{ -\frac{\beta_{i,j}(t+0.5) }{2N_{i,j}} \left(I_{i,j}(t) + I_{i,j}(t+1)\right) \right\}, \label{equation: approx sirdc model 1}\\ 
\frac{I_{i,j}(t+1)}{I_{i,j}(t)} \doteq&~\exp\left\{ \frac{\beta_{i,j}(t+0.5)}{2N_{i,j}} (S_{i,j}(t) + S_{i,j}(t+1)) - \gamma \right\}, \label{equation: approx sirdc model 2}\\
R_{i,j}(t+1) - R_{i,j}(t) \doteq&~\gamma \frac{I_{i,j}(t) + I_{i,j}(t+1)}{2} - \theta \frac{R_{i,j}(t) + R_{i,j}(t+1)}{2}, \label{equation: approx sirdc model 3}\\
D_{i,j}(t+1) - D_{i,j}(t) \doteq&~\delta \theta \frac{R_{i,j}(t) + R_{i,j}(t+1)}{2}, \label{equation: approx sirdc model 4}\\
C_{i,j}(t+1) - C_{i,j}(t) \doteq&~(1-\delta) \theta \frac{R_{i,j}(t) + R_{i,j}(t+1)}{2}. \label{equation: approx sirdc model 5}
\end{align}
Further by assuming the transmission rate parameter $\beta_{i,j}(t)$ is day-to-day invariant (i.e. a step function with step size 1){, based on} equations (\ref{equation: approx sirdc model 1}) and (\ref{equation: approx sirdc model 2}), we obtain $\beta_{i,j}(t+0.5)$ from $t = 1$ to $T_{i,j}-1$, iteratively, based on the sequence of susceptible individuals $\{S_{i,j}(t)\}_{t=1}^{T_{i,j}}$ and the initial number of active infectious individuals $I_{i,j}(1)$ described in algorithm \ref{algorithm:estimate_beta}. 


\begin{algorithm}[H]
 \KwData{$\{S_{i,j}(t)\}_{t=1}^{T_{i,j}}$, $I_{i,j}(1)$}
 \KwResult{$\{ \beta_{i,j}(t+0.5) \}_{t=1}^{T_{i,j}-1}$, $\{ I_{i,j}(t) \}_{t=1}^{T_{i,j}}$}
 $S_1 = S_{i,j}(1)$\;
 $S_2 = S_{i,j}(2)$\;
 $I_1 = I_{i,j}(1)$\;
 \For{$t=1$ to $(T_{i,j}-1)$}{
 
    $\beta_{i,j}\left(t+0.5\right) = \left\{\beta:  \frac{S_2}{S_1} - \exp\left\{-\frac{\beta I_1}{2 N_{i,j}} (1+\exp\{\frac{\beta}{2N_{i,j}} (S_1+S_2)-\gamma\})\right\} = 0 \right\}$ \;
    $I_{i,j}(t+1) = I_1 \exp\left\{ \frac{\beta_{i,j}(t+0.5)}{2N_{i,j}} (S_1 + S_2) - \gamma \right\}$ \;
    $S_1 = S_2$\;
    $S_2 = S_{i,j}(t+1)$\;
    $I_1 = I_{i,j}(t+1)$\;
 }
 \caption{Iterative approach for estimating transmission rate $\beta_{i,j}(t+0.5)$.}
 \label{algorithm:estimate_beta}
\end{algorithm}

After we get the number of active infective individuals ($I_{i,j}(t)$) on each day,  sequences of the resolving, deceased and recovered compartments can be solved subsequently following the same manner using equation (\ref{equation: approx sirdc model 3})-(\ref{equation: approx sirdc model 5}), after specifying their initial values. Expressing the time-dependent transmission rate by the number of susceptive and infective cases is the key to {integrating} death toll and testing data for estimation. 

In Extended Data Figure 1 and 2,  we demonstrate that {in order to solve the ODEs in the SIRDC model, our approach is more accurate and robust than the method $F\& J$ in\cite{fernandez2020estimating} under both simulated and real scenarios.} Other more accurate methods (such as the Runge-Kutta method) {can also} solve the ODEs of SIRDC model, {but} the time-dependent transmission rates {can} not easily {be} expressed as a function of the death toll and the number of active infectious individuals as the way they are in our solution.

\textbf{Estimation of the number of susceptible individuals}.  Note that we have 
 $S_{i,j}(t) + c_{i,j}^o(t) + c_{i,j}^u(t) = N_{i,j}$  for any $t$, where $c_{i,j}^o(t)$ and $c_{i,j}^u(t)$ are the number of cumulative observed confirmed cases and unobserved confirmed cases, respectively. Estimating the number of susceptible individuals is equivalent to estimating the number of unobserved confirmed cases $c_{i,j}^u(t)$, because the number of observed confirmed cases $c_{i,j}^o(t)$ and the population $N_{i,j}$ are known. Here we combine them with the positive test rates to estimate $c_{i,j}^u(t)$, as large positive test rates typically indicate a large number of unobserved confirmed cases.
  We assume that the total number of confirmed cases is equal to the observed confirmed cases, adjusted by the {state-level} test positive rate $p_{i}(t)$, a power parameter $\alpha_i$ and a weight parameter $\omega_i$, leading to the following formula of the susceptible population:
\begin{equation}
    S_{i,j}(t) = N_{i,j} - c_{i,j}^o(t) - c_{i,j}^u(t) = N_{i,j} - \frac{1}{\omega_{i,j}} \left\{ \mathbb{1}_{ \{t \ge 2\}} \sum_{s=2}^{t} (p_i (s))^{\alpha_i} \Delta c_{i,j}^o(s) + {(p_i (1))^{\alpha_i}} c_{i,j}^o(1) \right\}, 
    \label{equ:s_t}
\end{equation}
where $\Delta c_{i,j}^o(t)$ is the observed daily confirmed cases on day $t$, for $t=1,2,...,T_{i,j}$, $i=1,2,...,k$ and $j=1,2,...,n_{i}$. Since the positive test rates are only available at the state level, the power parameter $\alpha_i \in [0,2]$ is estimated by the state-level observations. According to  equation (\ref{equ:s_t}),  the time-invariant weight $\omega_{i,j}$ can be expressed below: 
\begin{equation}
    \omega_{i,j} =  \frac{{(p_i (1))^{\alpha_i}} c_{i,j}^o(1)}{I_{i,j}(1)+R_{i,j}(1)+D_{i,j}(1)+C_{i,j}(1)},
    \label{equation: function of omega}
\end{equation}
where $I_{i,j}(1)$, $R_{i,j}(1)$, $D_{i,j}(1)$ and  $C_{i,j}(1)$ are the number of active infectious, resolving, deceased and recovered cases on day $1$, respectively. 

\textbf{Estimation of initial values of infectious and resolving cases}. We define day $1$ of a county as {the more recent date between 21 March 2020 and the date that the county has $5$ observed confirmed cases for the first time.}
Since  all counties were at an early stage of the epidemic on the starting day, we let the initial value of {the}  death toll $D_{i,j}(1)$  be the observed death toll on {the} day $1$, and the initial value of the recovered cases  be $0$. This assumption is not likely going to {strongly} influence our analysis, as the number of recovered cases is only a negligible proportion of the susceptible individual on the starting day if not zero. 
The only parameters to estimate are the number of infectious individuals $I_{i,j}(1)$ and the number of resolving cases $R_{i,j}(1)$ on {the} day $1$ for {county $j$ from state $i$}, after the power parameter $\alpha_i$ is estimated using the state-level observations to minimize the same loss function below:  
\begin{align}
\begin{split}
(\hat I_{i,j}(1), \hat R_{i,j}(1))&= \argmin \sum_{t=1}^{T_{i,j}} \left(\frac{{D_{i,j}(t) - \hat{D}_{i,j}(t \mid I_{i,j}(1), R_{i,j}(1)) }}{T_{i,j}-t+1}\right)^2, \, s.t. \\
     0 &\le I_{i,j}(1)+R_{i,j}(1) \le U_{i,j},\,  I_{i,j}(1) \ge  0, \mbox{ and } R_{i,j}(1) \ge 0,
\end{split}
\label{equation:constrained loss function}
\end{align}
where the upper bound $U_{i,j}$ is chosen to guarantee the estimated number of the  susceptible cases $S_{i,j}(t)$ to be larger than $0$:
\begin{align*}
    U_{i,j} &= N_{i,j} \frac{{(p_i (1))^{\alpha_i}} c_{i,j}^o(1)}{\mathbb{1}_{\{T_{i,j} \ge 2\}} \sum_{s=2}^{T_{i,j}} (p_{i}(s))^{\alpha_{i}} \Delta c_{i,j}^o(s)  +{(p_i (1))^{\alpha_i}} c_{i,j}^o(1)} - (D_{i,j}(1) + C_{i,j}(1)),
\end{align*}
for $t=1,2,\dots, T_{i,j}$.  




After the initial values of infectious and resolving cases are estimated, we obtain the estimation of the susceptible cases  from equation (\ref{equ:s_t}), and the infectious cases and transmission rates on each date for each county from Algorithm 1. The resolving cases, deaths, and recovered {cases} can be derived subsequently from equation (\ref{equation: approx sirdc model 3})-(\ref{equation: approx sirdc model 5}), respectively. The estimated basic and effective reproduction rates can be derived by the fitted time-dependent transmission rate, and the estimated probability of contracting {SARS-CoV-2} for an individual can be computed based on transmission rate and number of infectious individuals for each county on each day. 

{\bf {Forecast and uncertainty assessment}}. Our method can also be used as a tool for forecasting  compartments (e.g., death toll), reproduction numbers, and the probability of contracting {SARS-CoV-2} at each county for a short period. We extrapolate the transmission rate based on Gaussian processes  implemented in {\tt RobustGaSP} R package \cite{gu2019robustgasp} with robust parameter estimation \cite{gu2018robust,gu2019jointly}. Based on the extrapolated transmission rates, the compartments can be solved iteratively based on equation (\ref{equation: approx sirdc model 1})-(\ref{equation: approx sirdc model 5}). 

We also found that the forecast will generally be improved by modeling  residuals between observed deaths and modeled deaths by a zero-mean Gaussian process (GP). 
One advantage of a GP model is the internal assessment of the uncertainty of the forecast from the predictive distribution, which is of crucial importance. The aggregated model that combines  the SIRDC model and the GP model for {county $j$ from state $i$} in the US is described as follows.
\begin{equation}
    D_{i, j}(t) = F_{i,j}(t) + z_{i,j}(t) + \epsilon_{i,j,t},
    \label{equ: model with GP}
\end{equation}
where $D_{i,j}(t)$ and $F_{i,j}(t)$ denote the observed death toll and estimated death toll via the SIRDC model, respectively;  The noise follows independently as a Gaussian distribution $\epsilon_{i,j,t} \sim N(0, \sigma_{i,j,0}^2)$ with variance parameter $\sigma_{i,j,0}^2$. The latent temporal process $z_{i,j}(t)$ is modeled by a zero-mean GP, meaning that for time points $\{1, 2, \dots, T_{i,j}\}$, $\mathbf{z}_{i,j} = \left(z_{i,j}(1), \dots, z_{i,j}(T_{i,j})\right)^T$ follows a multivariate normal distribution:
\begin{equation*}
    \mathbf{z}_{i,j} \sim \mbox{MN}(\mathbf{0}, \bm \Sigma_{i,j}),
\end{equation*}
where the $(l,m)$ entry of $\bm \Sigma_{i,j}$ is parameterized by a covariance function $\sigma^2_{i,j} K_{i,j}(l, m)$ for $1\leq l,m \leq T_{i,j}$. Here $\sigma^2_{i,j}$ is the variance parameter and $K_{i,j}(\cdot, \cdot)$ is a one-dimensional correlation function. We use the power exponential correlation function:
\begin{equation*}
    K_{i,j}(l, m) = \mbox{exp}\left\{ -\left(\frac{\mid l - m \mid }{ b_{i,j}} \right)^{a} \right\},
\end{equation*}
where $a$ is the roughness parameter fixed to be $1.9$ as in other studies \cite{bayarri2009using,gu2016parallel}, to avoid possible singularity in inversion of the covariance matrix using the Gaussian correlation ($a=2$), and $b_{i,j}$ is a range parameter for each county estimated from the data. 
We  define the nugget parameter $\eta_{i,j}=\sigma^2_{i,j,0}/\sigma^2_{i,j}$. The  range parameter $b_{i,j}$,  and the nugget parameter $\eta_{i,j}$ in equation (\ref{equ: model with GP}) are estimated based on the marginal posterior mode estimation using the {\sf rgasp} function in the package {\tt RobustGaSP} available on  CRAN \cite{gu2018robust}. 


Denote $\mathbf D_{i,j}=(D_{i,j}(1),...,D_{i,j}(T_{i,j}))^T$and  $\mathbf F_{i,j}=(F_{i,j}(1),...,F_{i,j}(T_{i,j}))^T$. 
After marginalizing out the variance parameter by the reference prior $p(\sigma^2_{i,j})\propto 1/\sigma^2_{i,j}$, for any $t^*$, the predictive distribution of $z_{i,j}({t}^*)$, conditional on the observations, range parameter $b_{i,j}$ and nugget parameter $\eta_{i,j}$, follows a non-central Student's t-distribution  with degrees of freedom $T_{i,j}$  \cite{gu2018robust}
\begin{equation}
z_{i,j}\left({t}^{*}\right) \mid \mathbf D_{i,j}, \mathbf F_{i,j}, b_{i,j}, \eta_{i,j} \sim \mathcal T \left(\hat{z}_{i,j}\left({t}^{*}\right), \hat{\sigma}^{2}_{i,j} \tilde K^{*}_{i,j}, T_{i,j}\right),
\label{equ: predictive distribution of GP}
\end{equation}
 where
\begin{equation*}
\begin{aligned}
\hat{z}_{i,j}\left({t}^{*}\right)=&  F_{i,j}(t^*)+ \mathbf{r}^{T}_{i,j}\left({t}^{*}\right) \mathbf{\tilde R}^{-1}_{i,j}(\mathbf D_{i,j}- \mathbf F_{i,j}), \\
\hat{\sigma}^{2}_{i,j}=&\frac{ (\mathbf D_{i,j}- \mathbf F_{i,j})^T \mathbf {\tilde R}_{i,j}^{-1} (\mathbf D_{i,j}- \mathbf F_{i,j})}{T_{i,j}},  \\
\tilde K^{*}_{i,j}=& K_{i,j}\left({t}^{*}, {t}^{*}\right)+\eta_{i,j}-\mathbf{r}^{T}_{i,j}\left({t}^{*}\right) \mathbf{\tilde R}^{-1}_{i,j} \mathbf{r}_{i,j}\left({t}^{*}\right),\\
\end{aligned}
\end{equation*}
with $\mathbf{\tilde R}_{i,j}=\mathbf R_{i,j}+\eta_{i,j}\mathbf I_{T_{i,j}}$, the $(l,m)$th term of $\mathbf R_{i,j}$ being $K_{i,j}(l,m)$ for $1\leq l,m\leq T_{i,j}$, and   $\mathbf{r}_{i,j}(t^*) = (K_{i,j}({t}^*, 1), \dots, K_{i,j}({t}^*, T_{i,j}))^{T}$,  by plugging in the estimated range parameter $b_{i,j}$ and nugget $\eta_{i,j}$. The predictive mean $\hat z_{i,j}(t^*)$ for forecasting the death toll of the $j$th county in the $i$th state at a future day $t^*$ and the predictive interval can be computed based on the  Student's $t$ distribution. An overview of the forecast algorithm and the numerical comparison of different approaches in forecast is given in the supplementary materials.



\section*{Data and code availability}
The datasets analysed in the current study are available in the CSSEGISandData repository, \url{https://github.com/CSSEGISandData/COVID-19} and COVID-19 data tracking project, \url{https://covidtracking.com/}.  The US maps are graphed based on publicly available R package {\tt urbnmapr}. The data and code used in this paper are
publicly available  (\url{https://github.com/HanmoLi/Robust-estimation-of-SARS-CoV-2-epidemic-in-US-counties}). 

\bibliography{COVID-19}



\section*{Acknowledgements}
This research is supported  by National Institute of Health 1R01DK130067 and by the UCSB Office of Research COVID-19 seed grant program. The authors thank the editorial board and two referees for their comments that substantially improved the article. The authors thank Siqi He for her contribution on the website of COVID-19 US Dashboard accompanied with this article. 
\section*{Author contributions statement}
H.L. analyzed data, developed the model, derived mathematical results, wrote computer code,  collected results and participated in manuscript writing. M.G. conceptualized the project, analyzed data, developed the model, derived mathematical results, wrote computer code,  analyzed results,  {and } led manuscript writing. 

\section*{Competing interests}
The authors declare no competing interests.

\newpage




%
%


{\centering
{\bf \Large Supplementary materials}
}

\beginsupplement

\quad 

\quad 

The supplementary materials contain three parts. In the first part, We discuss the details of  model parameter specification and conduct a {sensitivity analysis}. The forecast algorithm and numerical comparison of different approaches are introduced in the second part. The third part contains three videos for the county-level estimation of the daily PoC SARS-CoV-2 of a susceptible individual, the effective reproduction numbers, and the number of active infectious individuals from 21 March 2020 to 20 September 2020. 

\section{Model parameter specification and {sensitivity analysis}}
\label{section: sensitive analysis}

We discuss the choice of {the model parameters} and their sensitivity analysis. The following parameters of the SIRDC model were specified based on previous studies. 


\begin{itemize}
    \item  The death rate or the infection fatality ratio  ($\delta$) that measures the proportion of death among all infected individuals. We assume $\delta = 0.66\%$ following \cite{verity2020estimates}.

        \item The inverse of the number of days an infectious individual can transmit the COVID-19 ($\gamma$). The average time of a COVID-19 patient to transmit disease is assumed to be 5 days in \cite{davies2020effects}, indicating that $\gamma=0.2$. Another evidence comes from the study of {the} incubation period. 
         The latent period (exposed but not contagious) for COVID-19 is found to be 3.69 days on average \cite{li2020substantial} and the mean incubation period (time from infection to onset of symptoms) is 5.2 days \cite{li2020early}, meaning that {the} infectious period {is} around 1.5 days before the onset of symptom. The {diagnostic test} could take less than one day to up to {a} week. {We thus assume} 3.5 days to get the result of a {diagnostic test} on average. The total infectious period is around 5 {days}. 
        
        
    \item The inverse of the number of dates for resolving case to get resolved ($\theta$). According to the CDC report \cite{centers2020duration}, for mild and moderate symptom, the replication-competent virus has not been recovered after 10 days following symptom onset, indicating the individuals remains infectious no longer than 10 days after symptom onset. The infectious period could be as long as 20 days for patients with more severe illness from COVID-19 infection. Since a majority  of the COVID-19 infections are mild to moderate, we assume the infectious period to be 13.5 days, and  after reducing $3.5$ days from onset of the symptom to become resolving (after quarantine or {hospitalization}), it takes around 10 days for a resolving case to resolved on average. 
    
    
    
\end{itemize}

We conduct a {sensitivity analysis} to examine the change of the estimation in 4 different {configurations}. 


\begin{itemize}
    \item (\textbf{Configuration 1}) $(\gamma, \theta, \delta) = (0.2, 0.1, 0.0066)$, the  default parameter set.
    \item (\textbf{Configuration 2}) $(\gamma, \theta, \delta) = (0.14, 0.1, 0.0066)$. The average length of infectious period changes from 5 days to $\frac{1}{0.14}\approx 7$  days, whereas other parameters are held unchanged.
    \item (\textbf{Configuration 3}) $(\gamma, \theta, \delta) = (0.2, 0.067, 0.0066)$. The average length of resolving period changes from $10$ days to $\frac{1}{0.067} = 15$ days, whereas other parameters are held unchanged.
    \item (\textbf{Configuration 4}) $(\gamma, \theta, \delta) = (0.2, 0.1, 0.0075)$. The infection fatality ratio changes from $0.66\%$ to $0.75\%$, whereas other parameters are held unchanged.
\end{itemize}

\begin{figure}[h]
\centering

\includegraphics[height=0.5\textwidth,width=1\textwidth ]{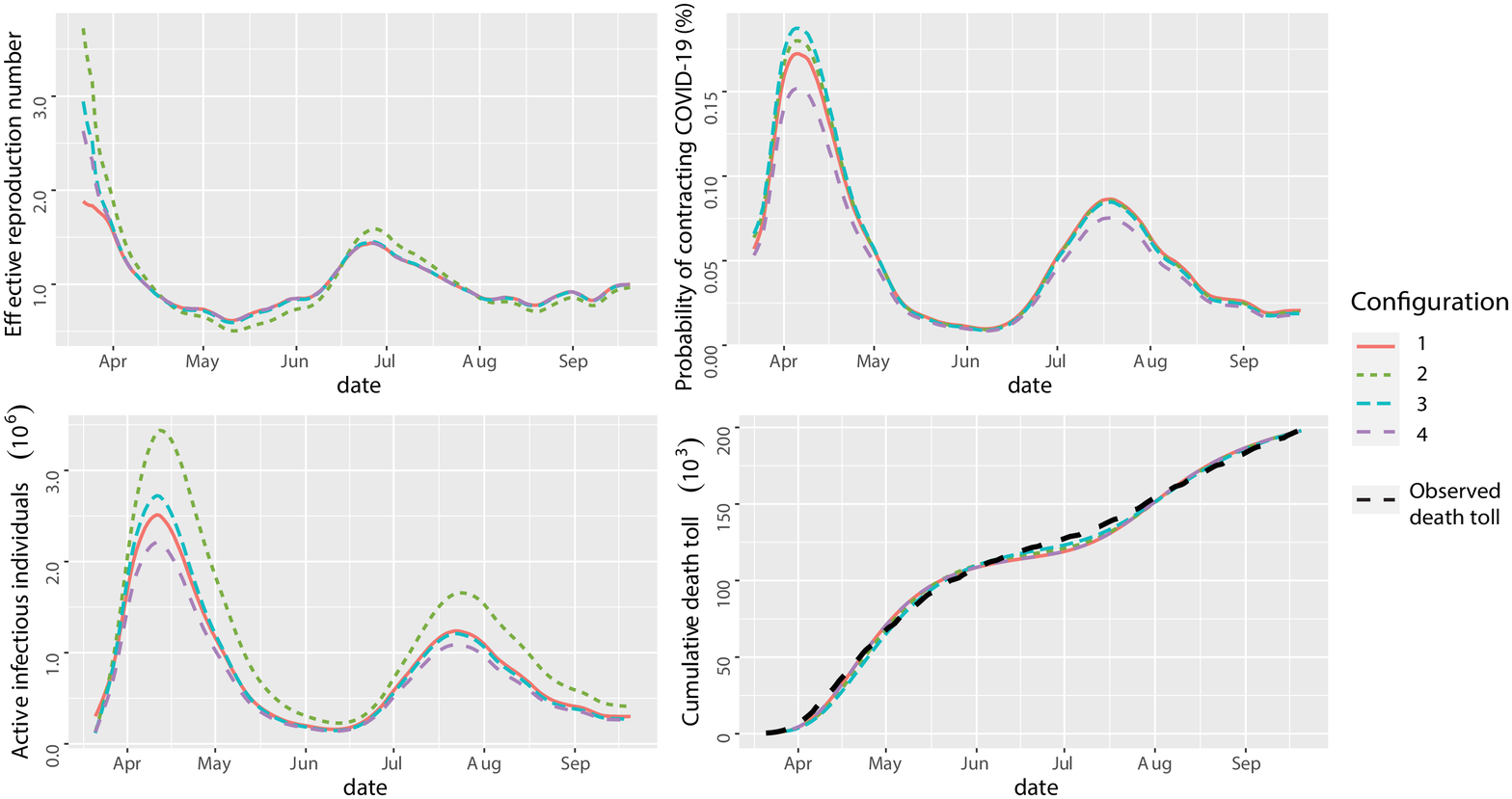}

\caption{{sensitivity analysis} for 4 configurations of the SIRDC model parameters. Part a-d {shows} the estimated effective reproduction number, PoC SARS-CoV-2, the number of active infectious individuals, and cumulative death toll, respectively.}
\label{fig:national_sensitive_analysis}
\end{figure}

After specifying the parameters $(\gamma, \theta, \delta)$, the transmission rate $\beta(t)$ can be obtained from algorithm 1. Figure \ref{fig:national_sensitive_analysis} gives result of the {sensitivity analysis}. First, we found the estimated death toll for 4 scenarios is almost the same (part d in Figure \ref{fig:national_sensitive_analysis}). Extending the infectious period  from 5 to 7 days (Configuration 2) increases the number of active infectious individuals  and effective reproduction number shown in part a and part c in Figure \ref{fig:national_sensitive_analysis}, respectively. Consequently, the peak of average daily PoC SARS-CoV-2 slightly increases in the first wave, whereas the scale of increase is  smaller than the change in the effective reproduction number and active infectious individuals. The  average daily PoC SARS-CoV-2  has almost no change {in} other periods, indicating that the length of the average infectious period has almost no influence of our estimation {on} PoC SARS-CoV-2 for most of the days.

Second, when the average length of {the} resolving period changes from  10 to 15 days, the peak of PoC SARS-CoV-2, effective reproduction number, and the number of active infectious individuals increases in the first wave, whereas these quantities  remain largely unchanged for the rest of the days (part a-c in Figure \ref{fig:national_sensitive_analysis}).  The result indicates that  the average length of {the} resolving period also barely affects the estimated characteristics of COVID-19 progression  {for most of the days}.

When the death rate increases from $0.66\%$ to $0.75\%$, the effective reproduction number seems to have almost no change (part a in Figure \ref{fig:national_sensitive_analysis}), whereas the PoC SARS-CoV-2 and the number of active infectious individuals  (figure \ref{fig:national_sensitive_analysis} part b-c) both reduce. This is because when the death rates {increase}, the estimated number of individuals infected decreases, as the death toll is observed (and thus fixed). The death rate is a key parameter to calibrate, and  studies of {the} prevalence of SARS-CoV-2 antibodies based on serology tests \cite{anand2020prevalence} can be used to estimate the death rate in each state. 

In conclusion,  parameter values of the average lengths of {the} infectious period and  {the} resolving period barely change the COVID-19 progression characteristics for most of the days, including the fitted death toll. On the other hand, we found that the number of active infectious individuals and the daily PoC SARS-CoV-2 depend critically on the death rate parameter. Further studies of prevalence would be useful for estimating the death rate {parameter} in different regions.  


\section{Algorithm of Forecast and numerical comparison} 
\begin{algorithm}[t]
 \caption{Ensemble forecast and uncertainty assessment. }
 
 \KwData{$\mathbf c_{i,j}^o$, $\mathbf D_{i,j}$, $\mathbf p_{i}$, $\mathbf c^o_i$, and $\mathbf D_i$.}
 \KwResult {Estimates  of county-level epidemiological compartments  $ \hat{\bm{\beta}}_{i,j} $, $ \hat{\mathbf{S}}_{i,j} $, $ \hat{\mathbf{I}}_{i,j}$, $ \hat{\mathbf{R}}_{i,j} $, $ \hat{\mathbf{C}}_{i,j}$, forecast $ \hat{\mathbf{D}}^*_{i,j} $, where $\hat{\mathbf{D}}^*_{i,j}:= \left(\hat{D}_{i,j}(T_{i,j}+1), \dots, \hat{D}_{i,j}(T_{i,j}+T^*)\right)^T$, and the  uncertainty assessment of the compartments.}
 
 \begin{description}
 \item \textbf{Step 1} Conduct a three-parameter constrained optimization treating state-level power parameter $\alpha_i$ unknown to minimize the loss function in equation (9) using $\mathbf p_{i}$, $\mathbf c^o_i$ and $\mathbf D_i$. 
 \item \textbf{Step 2} For each county, set initial values $I_{i,j}(1) = R_{i,j}(1) = 1,000$, $C_{i,j}(1) = 0$ and  $D_{i,j}(1)$ to be the observed death toll on day $1$. Find the optimized values of $I_{i,j}(1)$ and $R_{i,j}(1)$ to minimize equation (9). 
 \item \textbf{Step 3} Simulate $S=500$ samples of the  observed confirmed cases sampled from the predictive distribution of a GP model of the observed confirmed cases. For each sample, obtain the other compartments and time-dependent transmission rate by equation (1)-(5) and algorithm 1 using the estimate of the initial values.
 
\item  \textbf{Step 4} Extrapolate the time-dependent transmission rate parameters from a GP model for each sample and obtain $S=500$ samples of {the} output death toll of the SIRDC at the forecast period. 
\item \textbf{Step 5}  Sample the residuals  from the predictive distribution in Equation (11) in the main manuscript at the forecast period and obtain $S=500$ samples of the ensemble forecast for the death toll. Compute the mean for forecast and $95\%$ predictive interval to quantify  uncertainty of forecast. 

 \end{description}

 \label{algorithm:overview_algorithm}
\end{algorithm}

An overview of our algorithm for  forecast and  uncertainty assessment is given in algorithm \ref{algorithm:overview_algorithm}, where  inputs are the county-level observed cumulative number of confirmed cases $\mathbf c_{i,j}^o=(c_{i,j}^o(1),...,c_{i,j}^o(T_{i,j}))^T$, the county-level observed cumulative death toll $\mathbf D_{i,j}$, the state-level test positive rate $\mathbf p_{i}=(p_{i}(1),...,p_{i}(T_{i}))^T$, state-level confirmed cases $\mathbf c^o_i=(c^o_{i}(1),...,c^o_{i}({T_i}))^T$ and state-level death toll $\mathbf D_i=(D_{i}(1),...,D_{i}({T_i}))^T$.

To evaluate the performance of different approaches, we implement  7-day and 21-day forecasts on 2,277 US counties with {a} training period from 21 March 2020 to 20 September 2020, and with the forecast period starting from  21 September 2020. To compare the prediction performance of different methods,  we computed  the rooted mean square error (RMSE), {the} proportion of the observations covered in the $95\%$ predictive interval ($P_{CI}(95\%)$) and length  of the $95\%$ confidence interval ($L_{CI}(95\%)$), defined as follows:
\begin{align*}
    \textrm{RMSE} & = \sqrt{ \frac{\sum_{i=1}^{k} \sum_{j=1}^{n_i} \sum_{s \in \mathbf{t}^*} (\hat{D}_{i,j}(s) - D_{i,j}(s))^2  }{ \sum_{i=1}^{k} n_i T^*}} \\ 
    P_{CI}(95\%) & = \frac{1}{\sum_{i=1}^{k} n_i T^*} \sum_{i=1}^{k} \sum_{j=1}^{n_i} \sum_{s\in \mathbf{t}^*}   \mathbb{1}_{\{ D_{i,j}(s) \in CI_{i,j,s}(95\%) \}} \\
    L_{CI}(95\%) & = \frac{1}{\sum_{i=1}^{k} n_i T^*} \sum_{i=1}^{k} \sum_{j=1}^{n_i} \sum_{s\in \mathbf{t}^*} \textrm{length}\{ CI_{i,j,s} (95\%) \}
\end{align*}
where $\mathbf{t}^*:=(T_{i,j}+1, \dots, T_{i,j}+T^*)$, $T^*=7$ and $T^*=21$ for the 7-day forecast and 21-day forecast, respectively. A model with small RMSE, $P_{CI}(95\%)$ close to the nominal $95\%$ and small $L_{CI}(95\%)$ is precise  for forecast and uncertainty assessment.

\begin{table}[h]
\centering
\begin{tabular}{|c|cccc|}
\hline
\rowcolor{Gray}
{Prediction period}        & {Method}                  & {RMSE}  &  $P_{CI}(95\%)$ & $L_{CI}(95\%)$ \\ \hline
\multirow{4}{*}{7 days}  & SIRDC+GP                & \textbf{3.04}  & \textbf{$95.06\%$}       & 23.05     \\ & SIRDC                   & 4.12  &   /          &    /       \\
                         & GP without linear trend & 3.18  & $91.29\%$       & \textbf{4.82}      \\
                         & GP with linear trend    & 4.36  & $88.28\%$       & 5.51      \\
                         \hline
\multirow{4}{*}{21 days}  & SIRDC+GP                & \textbf{6.81}  & \textbf{$93.46\%$}       & 28.37  \\
& SIRDC                   & 7.79  &  /           &   /        \\
                         & GP without linear trend & 7.20  & $92.14\%$       & \textbf{11.74}     \\
                         & GP with linear trend    & 11.93 & $76.94\%$       & 10.18     \\
\hline
\end{tabular}
\caption{7-day and 21-day forecast {in} 2,277 US counties with training period from 21 March 2020 to 20 September 2020 and with prediction period starting from  21 September 2020. Four methods are compared.  Our proposed approach that combines the SIRDC model and a zero-mean GP to model the residuals is denoted as SIRDC+GP. Second, the death forecast by SIRDC model is denoted as SIRDC, which contains {Steps} 1 and 2 in the algorithm \ref{algorithm:overview_algorithm} and provides point projection of the death toll. Third, a GP with {a} constant mean  function is denoted as GP without linear trend, which  equivalently replaces the SIRDC model of a constant mean parameter estimated by the data for each county.
The fourth model,  denoted as GP with linear trend, is the same as the third method, except that the mean of GP contains {a} constant mean and a linear trend of time with two linear coefficient parameters estimated from the data. The best method under  each criterion is highlighted.}
\label{tab:comaprison_with_without_GP}
\end{table}

A comparison between our approach and the other three approaches {is} recorded in Table  \ref{tab:comaprison_with_without_GP}. Our approach (denoted in SIRDC+GP) has the lowest RMSE among 4 methods considered herein. Approximately $95\%$ of the held-out death toll are covered by the $95\%$ predictive interval by our approach, indicating our uncertainty assessment is accurate. Although other approaches produce {a} shorter length of the predictive interval, the number of held-out observations in the $95\%$ predictive interval is smaller than ours. Therefore, combining the SIRDC model and GP for modeling the residuals may  improve the predictive accuracy for forecasting COVID-19 associated death toll at US counties, compared to the one using the SIRDC model or the GP model alone. 

\clearpage

{\centering
{\bf \Large Extended data figures  for robust estimation of SARS-CoV-2 epidemic at US counties}
}

\captionsetup[figure]{name=Extended Data Figure}
\setcounter{figure}{0}

\begin{figure}[!htb]
\centering
\includegraphics[height=0.395\textwidth,width=1\textwidth ]{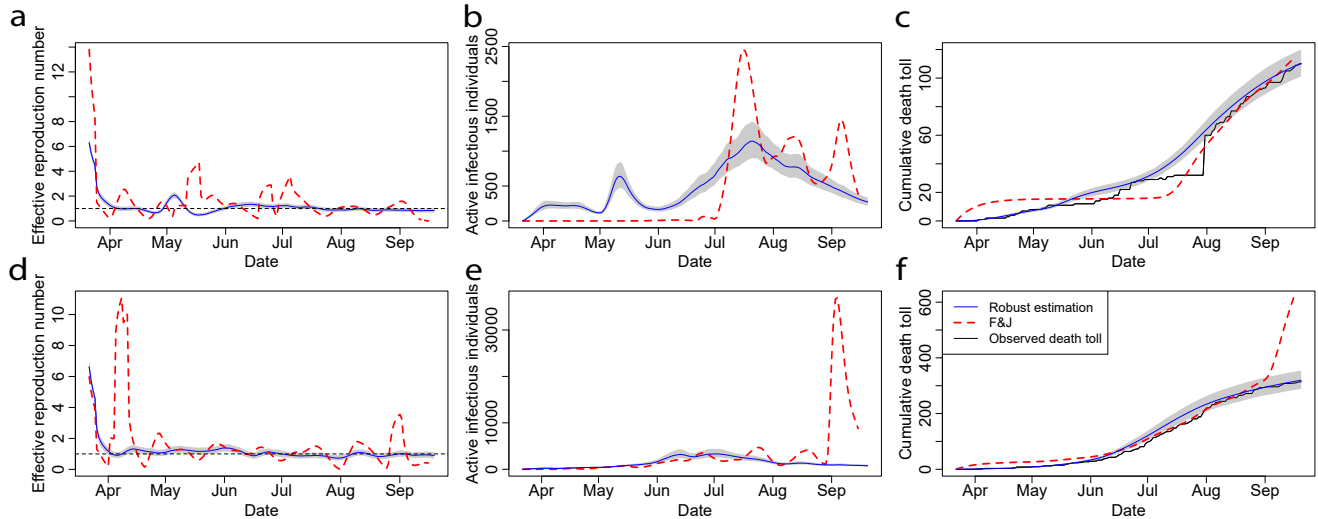}
   \caption{a-c, Comparisons between the estimation COVID-19 progression characteristics for Santa Barbara, CA as of 20 September 2020 by our algorithm 1 (blue solid curves) and the method $F\&J$ \cite{fernandez2020estimating} (red dash curves) . The shaded area represents $95\%$ confidence intervals. The black solid curve in part c is the observed cumulative death toll in Santa Barbara. d-f, Results for Imperial, CA as of 20 September 2020, which have the same interpretation as a-c. The transmission rate estimated from the method $F\&J$ is truncated to be within [0,10].}
    \label{fig:real_comparison_Euler}
\end{figure}
\vspace{-.2in}
\begin{figure}[!htb]
\centering
\includegraphics[height=0.385\textwidth,width=1\textwidth ]{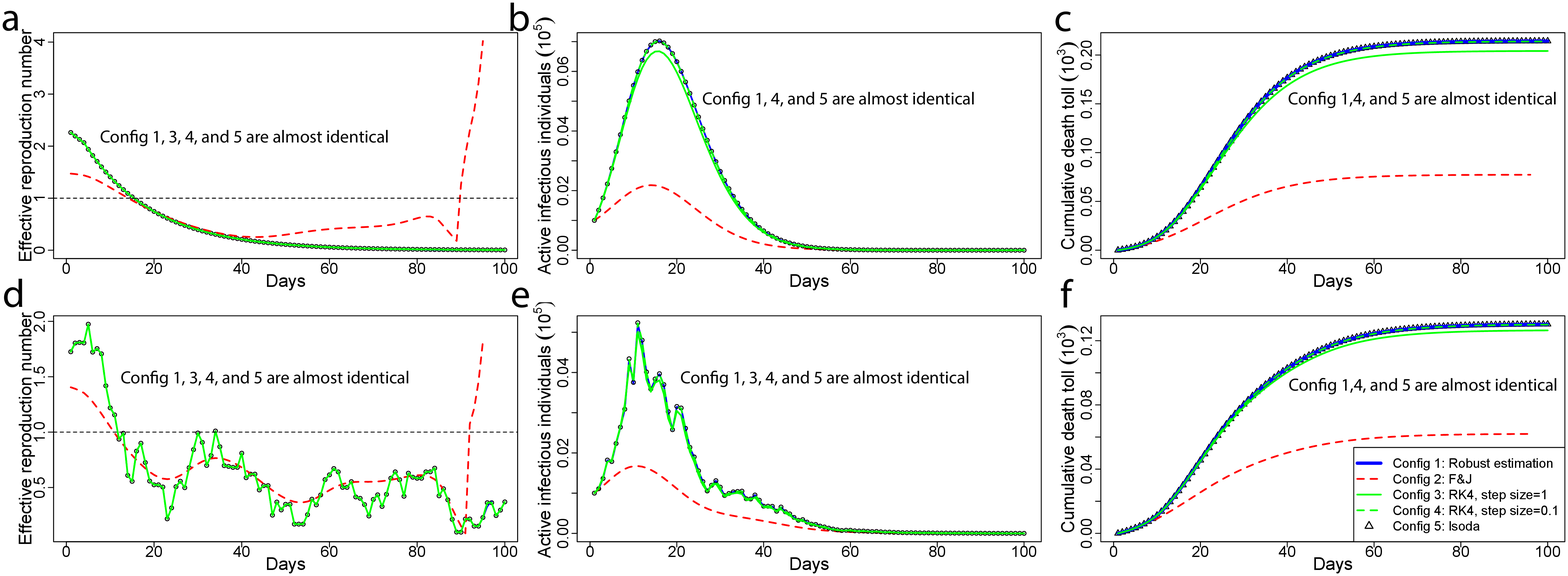}
   \caption{a-c, Simulated comparison with noise-free observations. The black circles are the solution of the ODEs of the SIRDC model via the default numerical solver Isoda in the function {\sf ode} in {\tt deSolve} {\tt R} package. The green solid and dash curves are the numerical solutions from Runge–Kutta method with the 4th order integration and step size being $1$ and $0.1$, respectively. The Blue solid curves are the robust estimation from algorithm 1 and red dash curves are the estimation in \cite{fernandez2020estimating}. In the simulation with noise-free observations, we let time duration be $T=100$ days, the population size $N=10^{7}$, the initial values of 5 compartments chosen as $(S(1), I(1), R(1), D(1), C(1))= (N-2000, 1000, 1000, 0, 0)$ and the transmission rate $\beta(t) = \mbox{exp}\left( -0.7 (\frac{9}{T-1} (t-1) + 1) \right)$, for $1\leq t \leq T$.  d-f, results of the simulation with noisy observations, which have the same interpretation as a-c.  In this simulation, we set the transmission rate $\beta(t) = \mbox{exp}\left( -0.7 (\frac{9}{T-1} (t-1) + 1) \right) + \epsilon$, for $1\leq t \leq T$ and $\epsilon \sim N(0, 0.04)$, and the other parameters are held the same as in the noise-free simulation. The transmission rates estimated from the method $F\&J$ are truncated to be within [0,10]. { The solution from our robust estimation approach, the Isoda and the Runge–Kutta method with the 4th order and step size being 0.1 overlap for both scenarios.} }
   \label{fig:simulated_comparison_Euler}
\end{figure}

\begin{figure}[!t]
\centering
\includegraphics[height=0.49\textwidth,width=1\textwidth ]{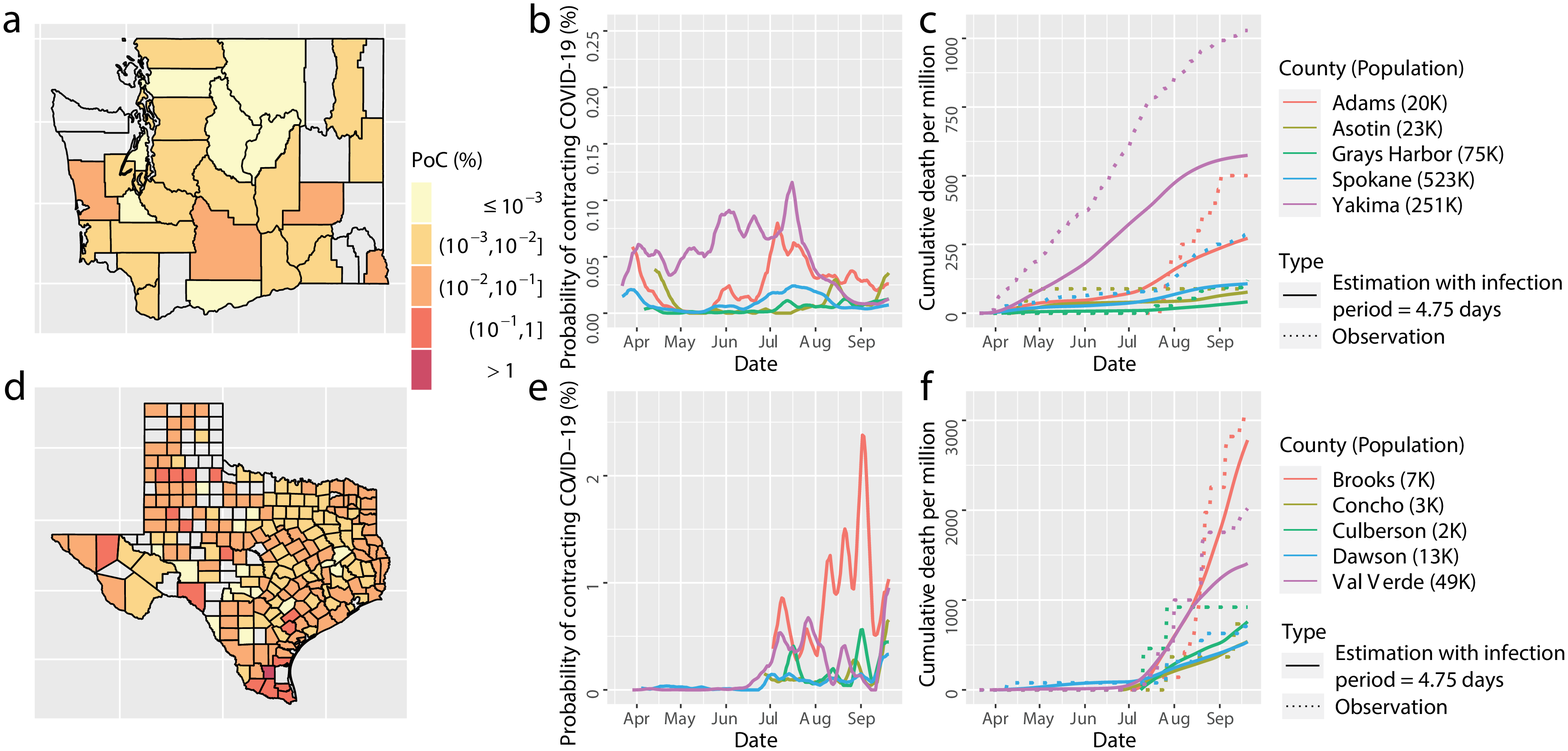}

\caption{a-f, the simulated results of COVID-19 progression in Washington (the first row) and in Texas (the second row) that have the same interpretation as a-f in Figure 3 with the infection period changed from 5 days, to 4.75 days, whereas other parameters are held the same. 
}
\label{fig:map_state_county_with_infection_period_4.75}
\end{figure}

\begin{figure}[!t]
\centering
\includegraphics[height=0.49\textwidth,width=1\textwidth ]{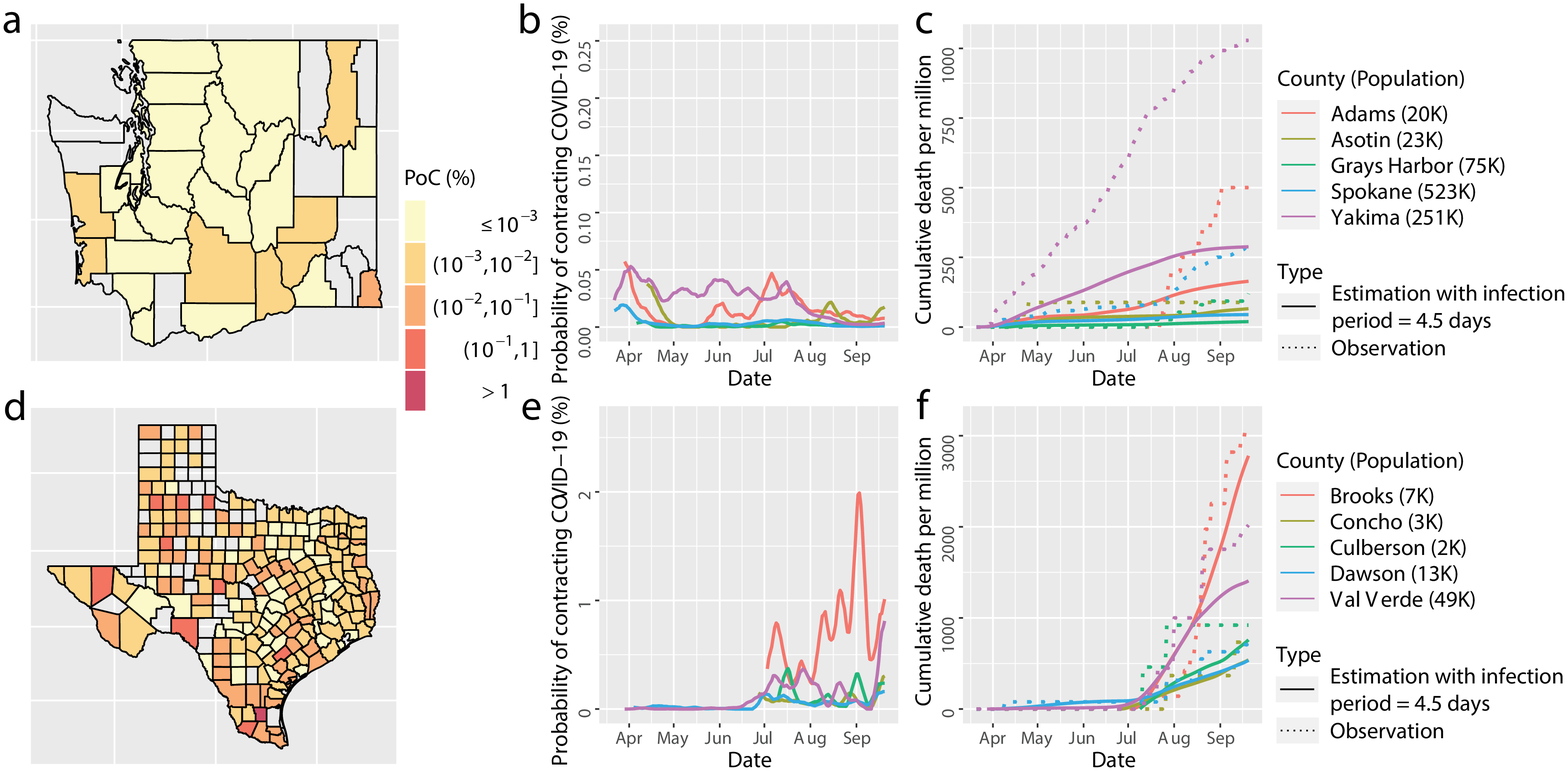}

\caption{a-f, the simulated results of COVID-19 progression characteristics in Washington (the first row) and in Texas (the second row) that have the same interpretation as a-f in Figure 3 with the infection period  changed from 5 days to 4.5 days, whereas other parameters are held the same. 
}
\label{fig:map_state_county_with_infection_period_4.5}
\end{figure}

\begin{figure}[!t]
\centering
\includegraphics[height=0.5\textwidth,width=1\textwidth ]{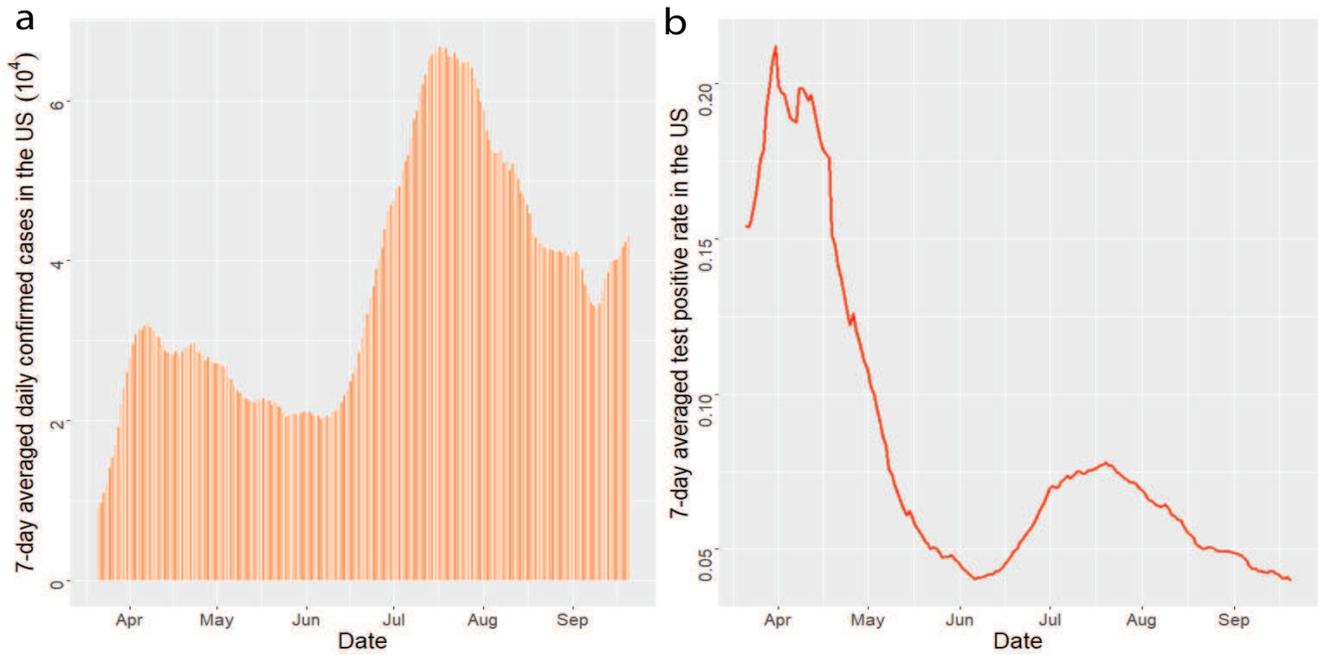}

\caption{a, the 7-day averaged daily confirmed cases in the US from 21 March 2020 to 20 September 2020. b, the 7-day averaged test positive rate in the US from 21 March 2020 to 20 September 2020.
}
\label{fig:confirmed_cases_positive_rate}
\end{figure}

\begin{figure}[h]
    \centering
    \includegraphics[height=0.45\textwidth,width=1\textwidth ]{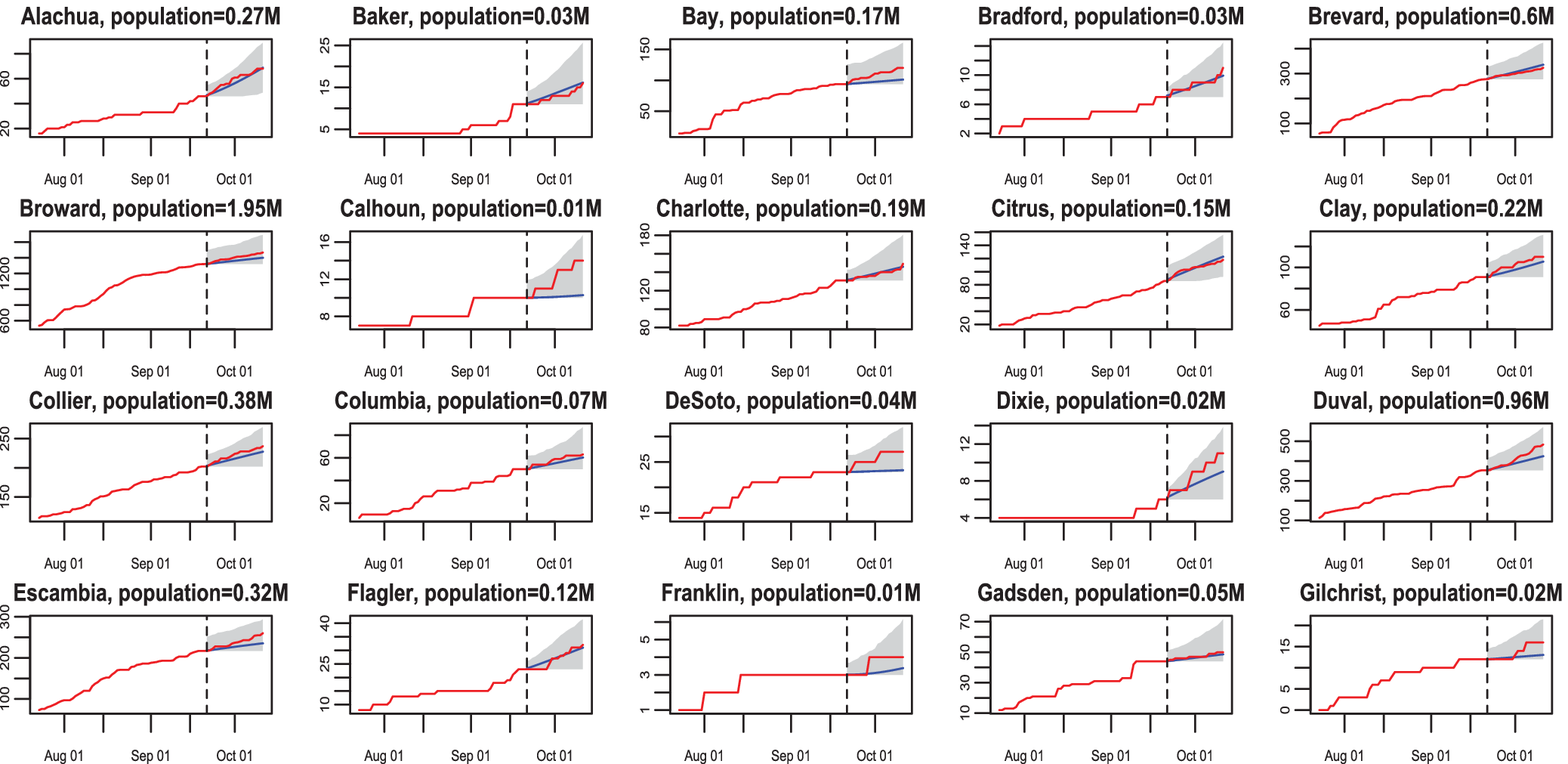}
    \caption{}
    \label{fig:90-day predictions in FL counties}
\end{figure} %
\begin{figure}[h]\ContinuedFloat
    \centering
    \includegraphics[height=1.1\textwidth,width=1\textwidth ]{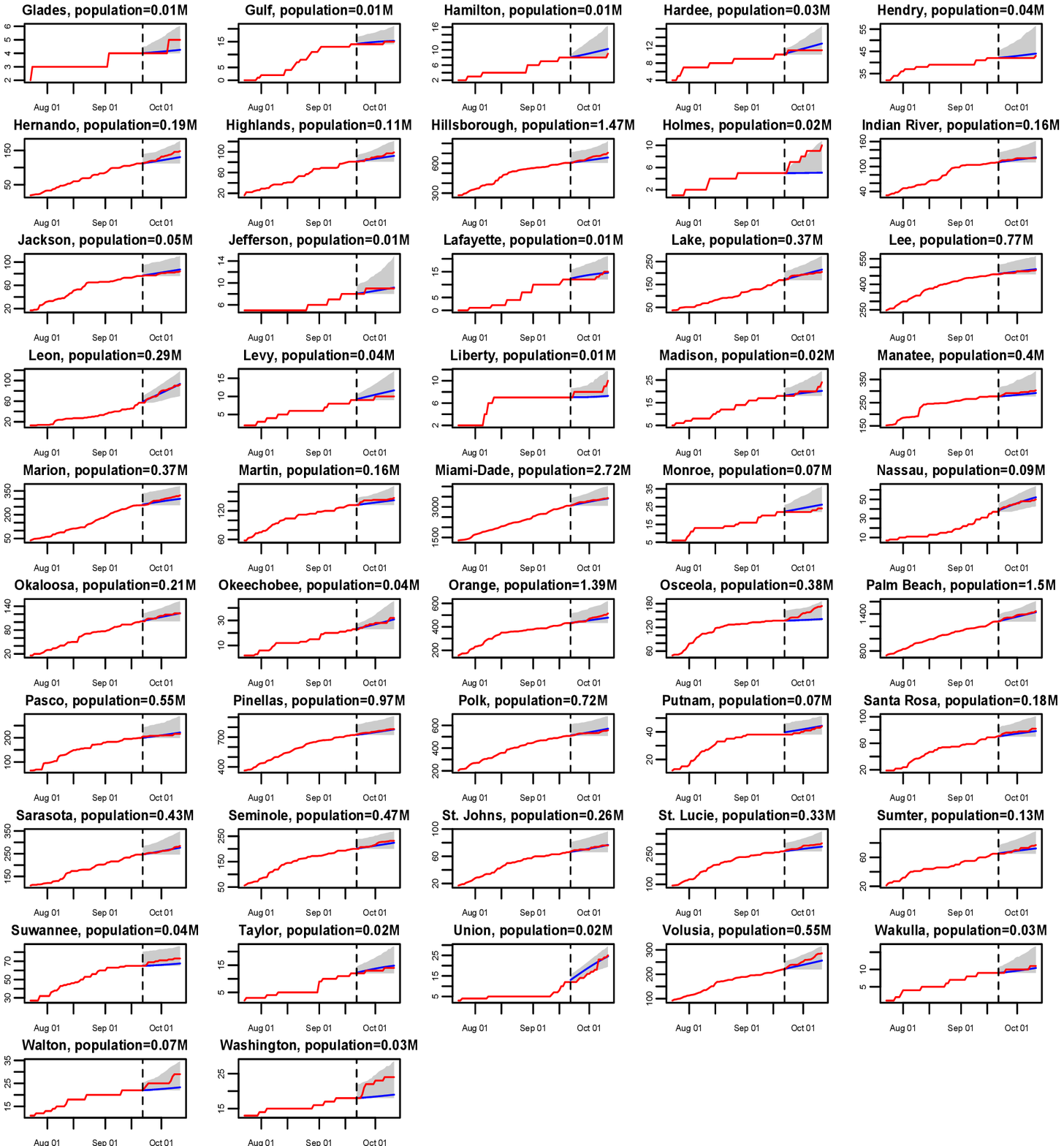}
    \caption{The 21-day forecast in 67 Florida counties with death toll no less than 2 as of 20 September 2020. The training period is from 21 March 2020 to 20 September 2020, whereas the forecast starts from 21 September 2020. The red curves are the cumulative observed death toll from 21 September 2020 to 11 October 2020 and the blue line indicates the forecast for the same period. The shaded area represents the $95\%$ predictive intervals of the forecast for each analyzed county in Florida.}
    \label{fig:90-day predictions in FL counties}
\end{figure}

\begin{figure}[t!]
    \centering
    \includegraphics[height=1.1\textwidth,width=1\textwidth ]{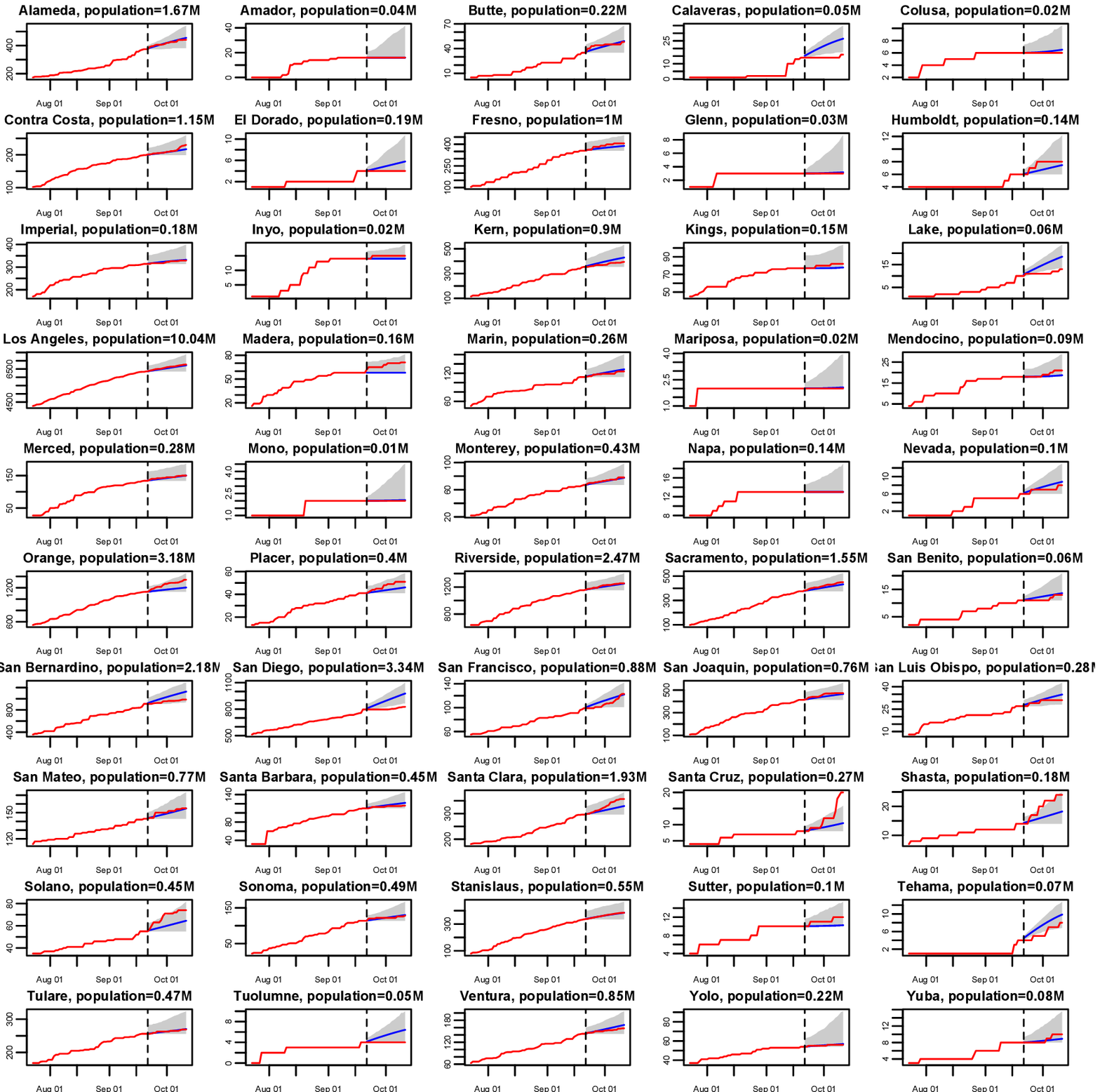}
    \caption{The 21-day forecast in 50 California counties with death toll no less than 2 as of 20 September 2020.  The training period is from 21 March 2020 to 20 September 2020, whereas the forecast starts from 21 September 2020. The red curves are the cumulative observed death toll from 21 September 2020 to 11 October 2020 and the blue line indicates the forecast for the same period. The shaded area represents the $95\%$ predictive intervals of the forecast for each analyzed county in California.}
    \label{fig:90-day predictions in CA counties}
\end{figure}

\begin{figure}[t!]
    \centering
    \includegraphics[height=0.49\textwidth,width=1\textwidth ]{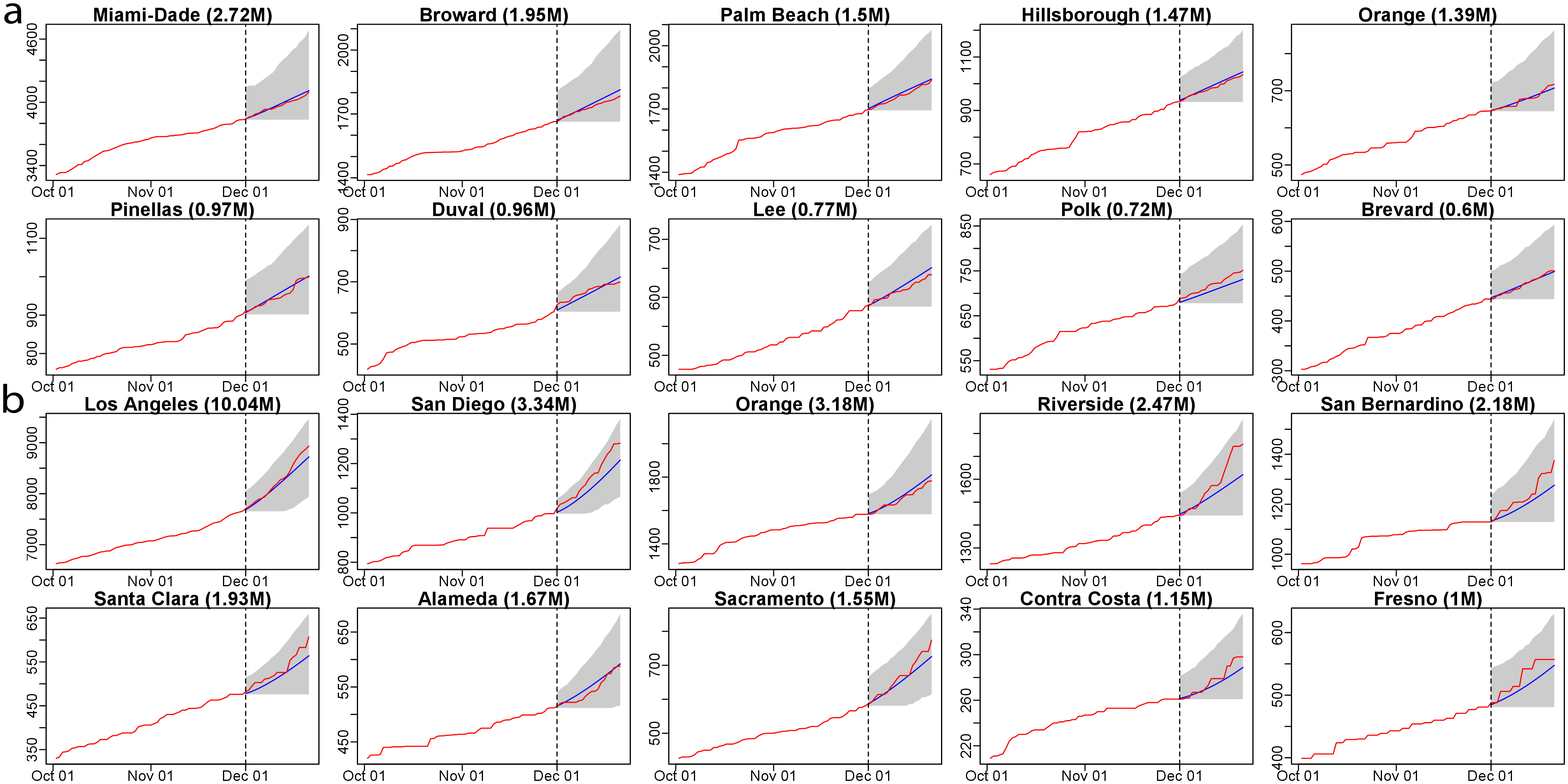}
    \caption{a, the 21-day forecast in 10 counties with the largest population in Florida.  The training period is from 21 March 2020 to 30 November 2020, whereas the forecast starts from 1 December 2020. The red curves are the cumulative observed death toll and the blue line indicates the forecast from 1 December 2020 to 21 December 2020. The shaded area represents the $95\%$ predictive intervals of the forecast for each analyzed county in Florida. The numbers in the parentheses are the populations in million for each county. b. the 21-day forecast in 10 counties with the largest population in California. The interpretations are the same as a.}
    \label{fig:21-day prediction in ten largest CA and FL counties}
\end{figure}

\begin{figure}[t!]
    \centering
    \includegraphics[height=0.49\textwidth,width=1\textwidth ]{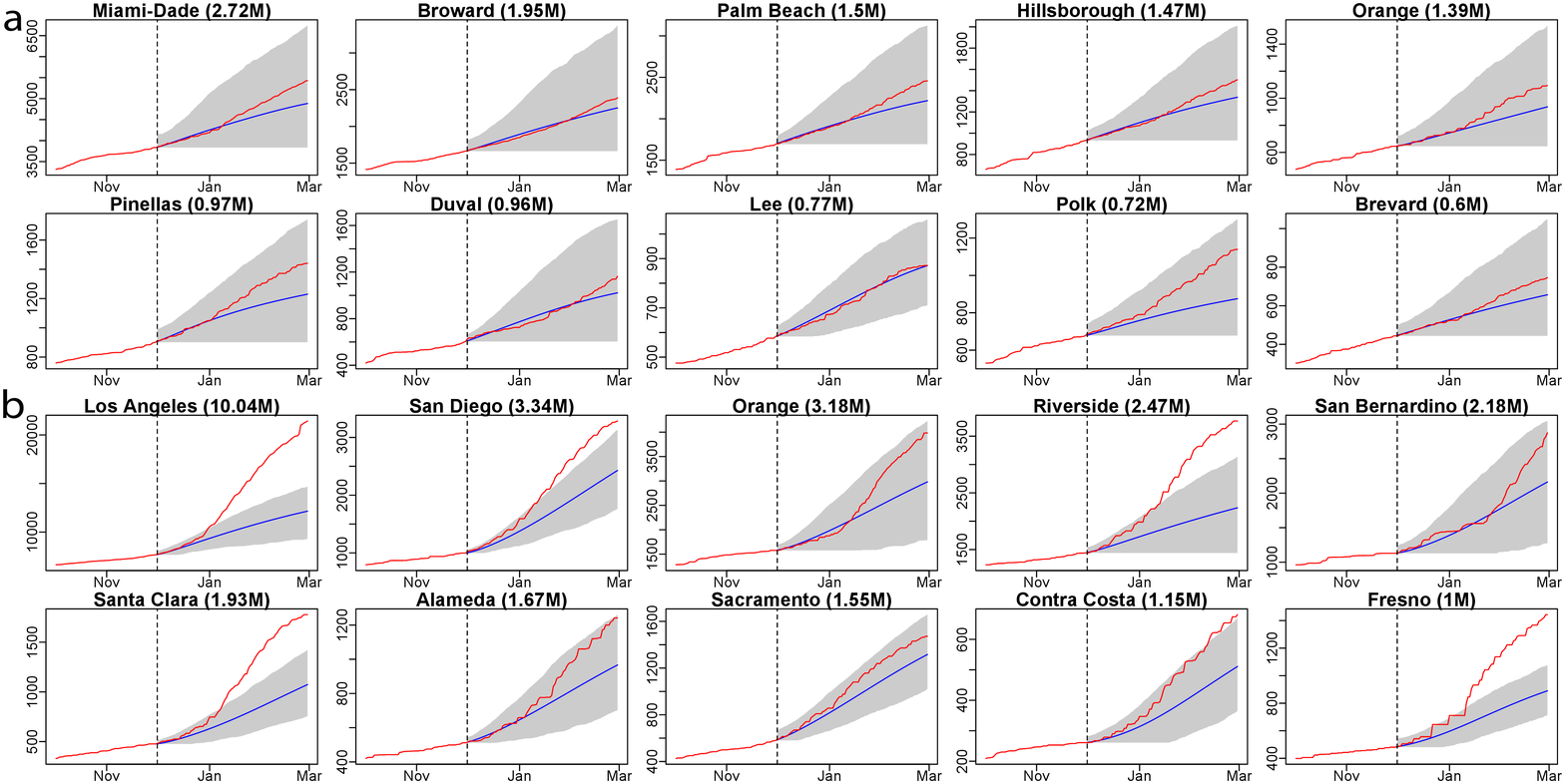}
    \caption{a, the 90-day forecast in 10 counties with the largest population in Florida.  The training period is from 21 March 2020 to 30 November 2020, whereas the forecast starts from 1 December 2020. The red curves are the cumulative observed death toll and the blue line indicates the forecast from 1 December 2020 to 21 December 2020. The shaded area represents the $95\%$ predictive intervals of the forecast for each analyzed county in Florida. The numbers in the parentheses are the populations in million for each county. b. the 90-day forecast in 10 counties with the largest population in California. The interpretations are the same as a.}
    \label{fig:90-day prediction in ten largest CA and FL counties}
\end{figure}

\end{document}